\documentclass[pdflatex,sn-vancouver-num]{sn-jnl}

\usepackage{graphicx}%
\usepackage{multirow}%
\usepackage{amsmath,amssymb,amsfonts}%
\usepackage{amsthm}%
\usepackage{mathrsfs}%
\usepackage[title]{appendix}%
\usepackage{xcolor}%
\usepackage{textcomp}%
\usepackage{manyfoot}%
\usepackage{booktabs}%
\usepackage{algorithm}%
\usepackage{algorithmicx}%
\usepackage{algpseudocode}%
\usepackage{listings}%

\usepackage{hyperref}
\usepackage{enumitem}
\usepackage{soul}
\usepackage{ulem}
\usepackage{caption}
\usepackage{subcaption}
\usepackage{adjustbox} 
\usepackage{collectbox}
\usepackage{fontawesome}
\usepackage[utf8]{inputenc}
\definecolor{darkyellow}{rgb}{0.85, 0.65, 0.13} 
\definecolor{red}{RGB}{249, 219, 218}
\definecolor{green}{RGB}{205, 245, 204}
\usepackage[table]{xcolor} 
\usepackage{float} 
\usepackage{tabularx}
\usepackage{makecell} 
\usepackage{array}
\usepackage{pifont}
\usepackage{rotating}
\usepackage{pdflscape}
\usepackage{color}

\IfFileExists{diffcolors.tex}{\input{diffcolors.tex}}{}

\theoremstyle{thmstyleone}%
\theoremstyle{thmstyletwo}%

\theoremstyle{thmstylethree}%

\begin{document}

\title[TraceLLM: Leveraging LLMs with Prompt Engineering for Enhanced Requirements Traceability]{TraceLLM: Leveraging Large Language Models with Prompt Engineering for Enhanced Requirements Traceability}

\author*[1,2]{Nouf Alturayeif}
\author[1,3]{Irfan Ahmad}
\author[4]{Jameleddine Hassine}

\affil[1]{Information and Computer Science Department, KFUPM, Dhahran, Saudi Arabia}
            
\affil[2]{Computing Department, Imam Abdulrahman Bin Faisal University, Dammam, Saudi Arabia}
            
\affil[3]{SDAIA--KFUPM Joint Research Center of Artificial Intelligence, KFUPM, Dhahran, Saudi Arabia}

\affil[4]{Computer Science Department, University of Quebec in Montreal, Montreal, Canada}

\abstract{
Requirements traceability, the process of establishing and maintaining relationships between requirements and various software development artifacts, is paramount for ensuring system integrity and fulfilling requirements throughout the Software Development Life Cycle (SDLC). Traditional methods, including manual and information retrieval models, are labor-intensive, error-prone, and limited by low precision. Recently, Large Language Models (LLMs) have demonstrated potential for supporting software engineering tasks through advanced language comprehension. However, a substantial gap exists in the systematic design and evaluation of prompts tailored to extract accurate trace links. 

This paper introduces TraceLLM, a systematic framework for enhancing requirements traceability through prompt engineering and demonstration selection. Our approach incorporates rigorous dataset splitting, iterative prompt refinement, enrichment with contextual roles and domain knowledge, and evaluation across zero- and few-shot settings. We assess prompt generalization and robustness using eight state-of-the-art LLMs on four benchmark datasets representing diverse domains (aerospace, healthcare) and artifact types (requirements, design elements, test cases, regulations). 

TraceLLM achieves state-of-the-art F2 scores, outperforming traditional IR baselines, fine-tuned models, and prior LLM-based methods. We also explore the impact of demonstration selection strategies, identifying label-aware, diversity-based sampling as particularly effective. Overall, our findings highlight that traceability performance depends not only on model capacity but also critically on the quality of prompt engineering. In addition, the achieved performance suggests that TraceLLM can support semi-automated traceability workflows in which candidate links are reviewed and validated by human analysts.}




\maketitle

\section{Introduction}

Requirements traceability refers to the ability to describe and follow the life of a requirement, in both forward and backward directions (i.e., from its origins, through its development and specification, to its subsequent deployment and use, and through periods of ongoing refinement and iteration)~\citep{gotel1994analysis,ramesh2002toward,alturayeif2025machine}. It enables tracking relationships between requirements and other software artifacts, including design elements, source code, and test cases.
It plays a pivotal role in ensuring that system requirements are consistently and correctly implemented, verified, and validated—especially in complex or safety-critical domains such as aerospace, healthcare, and finance~\citep{lin_traceability_2021}. In addition, effective traceability supports crucial software engineering tasks, including change impact analysis, compliance validation, and requirements verification~\citep{lin2022information}. Accurate traceability links help mitigate the risk of system failures by ensuring all requirements are adequately addressed throughout the Software Development Life-Cycle (SDLC)~\citep{lin_traceability_2021, rodriguez_prompts_2023,alturayeif2025machine}.

In practice, requirements traceability scenarios include Trace Creation, which establishes trace links during or after development; Trace Maintenance, which updates links as artifacts evolve; and Trace Integrity, which ensures link correctness and reliability through validation methods~\citep{cleland-huang_software_2014}. Trace Creation can be categorized into three main categories: Trace Link Completion (TLC), which fills in missing links between existing artifacts; Trace Link eXpansion (TLX), which links new artifacts to an established set as projects evolve; and Trace Link Generation (TLG), which creates trace links from scratch when no prior links exist~\citep{lyu2023systematic, lin2022enhancing, alturayeif2025machine}. Each task addresses unique traceability needs, ensuring that projects maintain comprehensive and accurate links across all development phases.

Despite its importance, requirements traceability remains a labor-intensive and error-prone process, particularly in large-scale projects where artifacts are numerous and complex~\citep{lin2022information, hayes2006advancing}. Traditional approaches, such as manual inspection or Information Retrieval (IR) techniques—including Vector Space Models (VSM)\citep{hayes2006advancing}, Latent Semantic Indexing (LSI)\citep{antoniol2002recovering}, and Latent Dirichlet Allocation (LDA)\citep{asuncion2011automated}—encounter challenges such as low precision and incomplete link generation~\citep{lin_traceability_2021}. Conventional Machine Learning (ML) approaches, such as Support Vector Machines (SVM) provide some enhancements~\citep{le2015rclinker, sun2017frlink, sun2017improving}, however, they heavily rely on handcrafted features like TF-IDF, which fail to capture intricate and context-dependent artifacts relationships~\citep{lan2023btlink}. Deep Learning (DL) approaches, such as LSTMs, offer better representation learning but are often limited by the need for large labeled datasets~\citep{zhu_enhancing_2022}. Transformer-based pretrained models like BERT and RoBERTa address semantic limitations by capturing contextual meaning through fine-tuning, but they still require labeled data and are restricted by task-specific adaptations, limiting their flexibility across diverse tasks and domains.

Large Language Models (LLMs) have recently demonstrated remarkable proficiency in in-context learning (ICL), where they make predictions based on a small set of demonstrations provided at inference time. This ability enables LLMs to adapt to new tasks without explicit fine-tuning, making them highly effective for various Natural Language Processing (NLP) applications such as question answering, text classification, and code generation. They possess substantial potential for automating software traceability by utilizing deep contextual embeddings to capture complex relationships between artifacts. This enables more accurate and semantically meaningful traceability links~\citep{rodriguez_prompts_2023}. However, realizing their promising potential, significant challenges remain in fully utilizing their capabilities. One of the most critical challenges is identifying effective prompting techniques that can be applied across various LLMs and software artifact types. Different models and datasets may necessitate distinct prompts for optimal results, and model behavior exhibits substantial variability (e.g., GPT~\citep{achiam2023gpt} versus Claude~\citep{askell2021general}), which complicates prompt design. Additionally, few-shot learning performance is highly sensitive to the selection of the demonstrations provided in the prompt. Different Demonstration Selection Strategies (DSSs)—such as random, similarity-based, or diversity-based sampling—can significantly influence trace link quality, but their effectiveness in the context of requirements traceability remains underexplored. Consequently, there is a critical need to explore both prompt design and DSS, and to evaluate their effectiveness across diverse LLMs in order to identify patterns and techniques that produce more accurate traceability outcomes.

This paper addresses the aforementioned challenges by presenting \textbf{TraceLLM}, an approach for prompt engineering with LLMs to enhance automated requirements traceability. To guide our investigation, this study addresses the following research questions in the context of automating requirements traceability between textual software artifacts:
\begin{itemize}
    \item \textbf{RQ1:} To what extent can prompt engineering improve the performance of LLMs in detecting trace links?
    \item \textbf{RQ2:} To what extent does a prompt designed for one LLM generalize to other LLMs, compared to LLM-specific prompts?
    \item \textbf{RQ3:} To what extent do demonstrations selection strategies influence few-shot learning performance?
    \item \textbf{RQ4:} To what extent does the proposed prompt-engineered LLM approach generalize across domains (e.g., healthcare, aerospace) and artifact types (e.g., requirements, test cases), compared to existing baselines and state-of-the-art traceability methods?
\end{itemize}

This pioneering study is the first to systematically explore and identify effective prompts for automated software traceability, evaluating their performance across a diverse range of LLMs. Specifically, this paper makes the following key contributions:
\begin{itemize}
    \item Establish an LLM-based traceability methodology with standardized dataset splitting, prompt design, and evaluation. This ensures consistency, transparency, and replicability.
    \item Design and evaluate tailored prompts for detecting trace links between textual artifacts using the CM1 dataset as a case study. In addition, experiment with multiple DSSs to assess their impact on few-shot learning performance.
    \item Evaluate the designed prompt across eight advanced LLMs and four datasets of different artifacts, benchmarking against baselines and state-of-the-art methods to demonstrate improvements in automated requirements traceability.
    \item Provide a replication package\footnote{\url{https://anonymous.4open.science/r/TraceLLM-4C72/}}, including our code and experimental setup, to facilitate further research in LLM-based automated requirements traceability.
\end{itemize}

The paper is organized as follows: Section~\ref{sec:related_work} covers related work, Section~\ref{sec:Research_Methodology} introduces the proposed methodology (TraceLLM), Section~\ref{sec:DSSs} describes the most commonly used DSSs examined in this study. Section~\ref{sec:exp_setup} details the experimental setup, Section~\ref{sec:results} presents results. Section~\ref{sec:TTV} discusses threats to validity and Section~\ref{sec:conclusion} concludes with future work directions.

\section{Related Work}\label{sec:related_work}

Automated requirements traceability has been explored through a variety of paradigms, ranging from traditional IR to advanced language models. This section categorizes prior research into five thematic groups: IR methods, conventional ML models, DL-based approaches, pretrained language models, and in-context learning using LLMs. Each category is discussed below, followed by a summary table and a discussion of how this study extends the state-of-the-art.

\textbf{Information Retrieval (IR).} IR techniques were among the earliest to support traceability tasks by computing textual similarity between software artifacts~\citep{antoniol2002recovering, hayes2006advancing, hayes2003improving, lucia2007recovering, oliveto2010equivalence}. Common models include the VSM~\citep{hayes2006advancing}, LSI~\citep{antoniol2002recovering}, and LDA~\citep{asuncion2011automated}. These models are simple and domain-agnostic, however, their reliance on exact matches and simple term frequencies limits their ability to capture deeper semantic relationships that are essential for effective traceability.

\textbf{Conventional Machine Learning (ML).}
Conventional ML methods, such as SVM and Naive Bayes (NB), advanced traceability by learning patterns from training data and were widely utilized in numerous automated traceability studies~\citep{workneh2023machine, wang2023empirical, van2023effectiveness, moran2020improving, mills2018automatic, rath2018traceability, mills2017automating}. These models are typically trained on features derived from term frequency–inverse document frequency (TF-IDF), similarity scores, and document statistics \citep{mills2018automatic, moran2020improving, rath2018traceability}.

\citet{zhao_improved_2017} introduced WELR, a learning-to-rank approach that integrates word embeddings with a weighting strategy to improve ranking precision. \citet{tian_adapting_2018} similarly adapted word embeddings for traceability, focusing on handling out-of-vocabulary terms and combining similarity metrics with an ML ranker. TRAIL \citep{mills2018automatic} further advanced ML-based traceability by leveraging historical trace links to train a classifier capable of predicting new and updated links. \citet{du_automatic_2020} proposed an active learning variant to reduce annotation costs. While effective, these models require feature engineering and do not capture deeper semantics and complex relationships between software artifacts.

\textbf{Deep Learning (DL).} DL models, such as Long Short-Term Memory (LSTM) networks, have shown promise by automatically learning semantic representations from textual artifacts. \citet{chen2019enhancing} introduced S2Trace, an unsupervised approach that uses sequential pattern mining and Doc2Vec embeddings to model both the semantics and the ordering of terms in requirements. Unlike previous methods, S2Trace requires no labeled data and achieved superior performance over traditional baselines. Although early DL approaches improved generalization, their reliance on large labeled datasets poses challenges in traceability tasks where data is often limited or imbalanced.

\textbf{Pretrained Language Models.} 
Pretrained language models such as BERT and RoBERTa are based on the transformer encoder architecture and are typically trained on large corpora using masked language modeling objectives. These models are moderately sized, ranging from 110M (BERT-base) to 350M (BERT-large) parameters, and are commonly fine-tuned on downstream tasks with limited amounts of labeled data.

In the context of traceability, fine-tuning transformer models offer strong performance by learning task-specific patterns while benefiting from general-purpose language understanding. For example, NLTrace~\citep{lin2022enhancing} leveraged transfer learning with BERT to support trace link generation, completion, and expansion. The model was fine-tuned on GitHub datasets and demonstrated substantial gains over classical IR and ML techniques, particularly in low-data scenarios. Similarly, Kashif~\citep{etezadi2025classification} fine-tuned sentence transformers (typically based on BERT or RoBERTa) for requirements-to-regulation traceability. However, the model’s performance dropped significantly when applied to complex regulatory documents, suggesting limited generalization. While these encoder-only models are computationally efficient and effective for many supervised tasks, they require labeled data for fine-tuning and typically involve task-specific training, raising barriers to scalability and generalizability.

\textbf{In-Context Learning (ICL) with Large Language Models (LLMs).}

In contrast to encoder-only models like BERT, LLMs such as GPT-3, GPT-4, Claude, and LLaMA are decoder-only transformers with vastly larger number of parameters, typically ranging from 6B to 175B+ parameters. These models support ICL, where predictions are conditioned on demonstrations and instructions embedded directly in the prompt, without updating model weights. This approach has become increasingly popular in traceability due to its low setup cost and strong few-shot capabilities. 

While several studies have explored the use of prompt engineering in software engineering tasks, few have proposed general-purpose prompting strategies for automated traceability~\citep{rodriguez_prompts_2023, hey2025requirements}. Some focused on specific artifact types, such as requirements-to-regulations~\citep{etezadi2025classification, masoudifard2024leveraging}, requirements-to-goal models~\citep{hassine2024llm}, system-to-user requirements~\citep{niu2025tvr}, and architecture documentation-to-source code~\citep{fuchss2025enabling}, requirements-to-source code~\citep{fuchss2025lissa}. Other studies addressed related but distinct tasks, including high-level to low-level requirements coverage~\citep{preda2024supporting} and impact of requirements smells on LLMs performance~\citep{vogelsang2025impact}. In the following, we briefly review the studies most closely related to ours and highlight the specific research gaps our work addresses.

\citet{rodriguez_prompts_2023} introduced a preliminary study on leveraging prompts to enhance the automation of traceability links in complex software projects. Their approach demonstrated the potential of prompt engineering to adjust the language model's understanding in specific traceability tasks, resulting in significant improvements in recall while encountering challenges in balancing precision. On the CM1 dataset, they achieved a precision of 0.38, recall of 0.85, and F2 Score of 0.43. However, it is important to note that their study was preliminary, utilizing only a small portion of the dataset and reporting results on the same partition used for prompt design, which introduces data leakage and limits the generalizability of their findings. In contrast, our study establishes a rigorous data-splitting method that avoids leakage and evaluates prompts on a distinct test set, ensuring an unbiased performance assessment. Furthermore, we utilize the complete datasets with different types of textual artifacts, providing a more comprehensive evaluation of prompt performance and enhancing the robustness of our findings. 

\citet{hey2025requirements} proposed a Retrieval-Augmented Generation (RAG) approach for inter-requirements traceability, specifically targeting links between high-level and low-level requirements as well as requirements and regulations. Their RAG component is a retrieval step that filters out unrelated candidate pairs based on embedding similarity before prompting the LLM. The study showed that Chain-of-Thought (CoT) prompting outperforms simpler classification-based prompts. It also demonstrated that open-source LLMs (e.g., LLaMA) can achieve comparable results to proprietary models (e.g., GPT-4o). However, their evaluation did not isolate the effect of the retrieval step by comparing performance with and without RAG. Additionally, their study did not explore few-shot learning or prompt engineering strategies and was limited to inter-requirements traceability. In contrast, our work focuses on prompt design and DSSs, evaluates both zero-shot and few-shot prompting, and investigates generalizability across different textual artifact types and domains.

\citet{masoudifard2024leveraging} and \citet{etezadi2025classification} proposed LLM-based approaches specific to tracing requirements to regulations. \citet{masoudifard2024leveraging} introduced a RAG-based approach that integrates graph-based retrieval from regulatory texts with prompt engineering techniques on GPT-4o, including CoT and Tree of Thought (ToT) techniques. Their method significantly improves performance over baseline RAG techniques, particularly in identifying non-compliant requirements. 
However, while effective, this approach comes with higher computational costs and increased complexity, making it more challenging to implement across diverse contexts. \citet{etezadi2025classification} employed GPT-4o using the RICE prompting framework, which includes Role, Instruction, Context, Constraints, and Examples (i.e., demonstrations). Their findings showed that the prompt-based approach outperformed the fine-tuned model. In contrast, our study takes a broader perspective by focusing on the process of prompt engineering and offering a comprehensive comparison of various well-known LLMs across different datasets. 

\citet{hassine2024llm} employed a Zero-Shot approach utilizing GPT-3.5-turbo (a legacy version of GPT) to trace security requirements to goal models. In contrast, our study builds upon these efforts by employing both zero-shot and few-shot settings and utilizing eight recent LLMs on diverse datasets. Additionally, we conducted an extensive process of prompt design and refinement. 

LLMs bring scalability and generalization capabilities to traceability tasks, but their effectiveness is highly dependent on prompt quality, demonstrations selection, and robust evaluation. Table~\ref{tab:llm_comparison} shows a comparison of the related work and this study. There are several limitations in the existing literature. Most notably, prior work tends to focus on either a single model or a specific domain, often lacking rigorous prompt engineering or comprehensive evaluation across models and tasks. Additionally, the influence of DSSs on few-shot learning performance remains largely unexplored. This paper addresses these gaps by proposing a systematic evaluation methodology (TraceLLM) that handles dataset splitting strategies to represent different tractability scenarios, investigates prompt design, and explores DSSs. In addition, this study assesses the generalization of prompts across different advanced LLMs and multiple datasets.

\begin{sidewaystable}[]
\centering
\small
\caption{Comparison of related LLM-based traceability studies.}
\begin{tabular}{p{2cm}lllll}
\toprule
\textbf{Study} & \textbf{Model(s)} & \begin{tabular}[c]{@{}l@{}}\textbf{Prompting}\\ \textbf{Strategy}\end{tabular} & \textbf{Dataset(s)} & \textbf{Artifact Types} & \textbf{Pros/Cons} \\
\midrule

\citet{rodriguez_prompts_2023} &
Claude instant v1 &
Zero-shot  &
\begin{tabular}[c]{@{}l@{}}CM1,\\ Dronology,\\ iTrust\end{tabular} &
\begin{tabular}[c]{@{}l@{}}Requirements to design,\\ requirements to code,\\ design to code\end{tabular} &
\begin{tabular}[c]{@{}l@{}}
Explored prompt variants,\\
data leakage, no DSSs, no\\
cross-model or cross-\\
domain analysis.\end{tabular} \\

\midrule

\citet{hey2025requirements}
& \begin{tabular}[c]{@{}l@{}}GPT-4o-Mini,\\ GPT-4o,\\ LLaMA 3.1 8B,\\ CodeLLaMA 13B\end{tabular}
& \begin{tabular}[c]{@{}l@{}}Zero-shot,\\ CoT\end{tabular} &
\begin{tabular}[c]{@{}l@{}}CM1,\\ Dronology,\\ CCHIT,\\ MODIS,\\ GANNT,\\ WARC\end{tabular} &
\begin{tabular}[c]{@{}l@{}}Requirements to design,\\ requirements to\\ regulations\end{tabular} &
\begin{tabular}[c]{@{}l@{}}
Used CoT, compared open-\\
source to proprietary LLMs,\\ 
no prompt engineering and\\ DSS analysis.\end{tabular} \\

\midrule

\citet{etezadi2025classification}&
GPT-4o &
5-shots &
\begin{tabular}[c]{@{}l@{}}KeePass,\\ WASP,\\ Datahub,\\ ScrumAlliance\end{tabular} &
\begin{tabular}[c]{@{}l@{}}Requirements to\\ Regulations\end{tabular} &
\begin{tabular}[c]{@{}l@{}}
Structured prompting with\\
demonstrations and\\
rationale, specific to\\
requirements to regulations,\\
one LLM.\end{tabular}\\

\midrule

\citet{masoudifard2024leveraging} &
GPT-4o &
\begin{tabular}[c]{@{}l@{}}Few-shot,\\ CoT, ToT\end{tabular} &
Private &
\begin{tabular}[c]{@{}l@{}}Requirements to\\ Regulations\end{tabular} &
\begin{tabular}[c]{@{}l@{}}
Heavy and complex,\\
regulatory focus, no LLM\\
diversity or generalization\\
study.\end{tabular} \\

\midrule

\citet{hassine2024llm}&
GPT-3.5 &
Zero-shot &
Private &
\begin{tabular}[c]{@{}l@{}}Security requirements\\ to goal models\end{tabular} &
\begin{tabular}[c]{@{}l@{}}
Focused on security domain,\\
did not report prompt\\
tuning, no models\\
comparison. \end{tabular} \\

\midrule

\begin{tabular}[c]{@{}l@{}}\textbf{This Study}\\ \textbf{(TraceLLM)}\end{tabular} &
\begin{tabular}[c]{@{}l@{}}GPT-4o,\\ GPT-4o-Mini,\\ Claude 3.5 Haiku,\\  Claude 3.5 Sonnet,\\ Gemini 1.5 Flash,\\ Gemini 1.5 Pro,\\ LLaMA 3.1 8B,\\ LLaMA 3.1 70B\end{tabular} &
\begin{tabular}[c]{@{}l@{}}Zero and\\ few-shot\\ (random\\ and DSSs)\end{tabular} &
\begin{tabular}[c]{@{}l@{}}CM1,\\ EasyClinic (UC-TC),\\ EasyClinic (UC-ID),\\ CCHIT\end{tabular} &
\begin{tabular}[c]{@{}l@{}}Requirements to design,\\ use cases to test cases,\\ use cases to interaction\\ diagrams, requirements\\ to regulations\end{tabular} &
\begin{tabular}[c]{@{}l@{}}
Evaluates 8 LLMs, explore\\
different DSSs, cross-domain\\
and artifacts generalization,\\
rigorous data splitting.\end{tabular} \\

\bottomrule
\end{tabular}
\label{tab:llm_comparison}
\end{sidewaystable}

\section{LLM-Based Traceability Methodology}\label{sec:Research_Methodology}
\begin{figure*}[htb!]
    \centering
    \includegraphics[width=\linewidth]{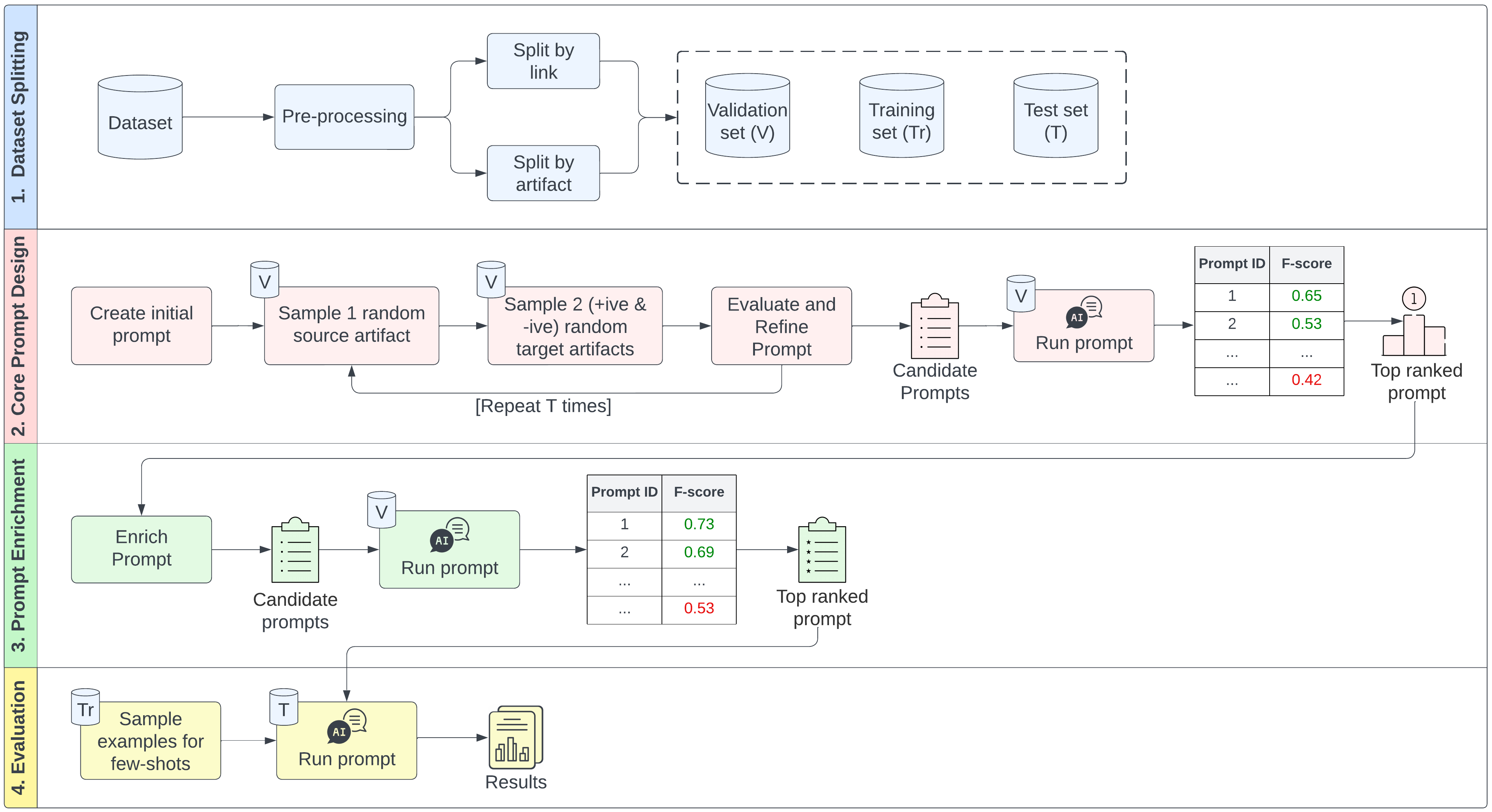}
    \caption{LLM-based traceability methodology (TraceLLM)}
    \label{fig:Methodology}
\end{figure*}

In this study, we developed a four-stage methodology for iteratively designing and evaluating prompts for software traceability using LLMs (TraceLLM). The methodology involves dataset splitting, core prompt design, prompt enrichment, and evaluation. TraceLLM provides a structured guide for prompt engineering, a process that is inherently iterative and human-in-the-loop~\citep{schulhoff2024prompt}, for which no principled method currently exists to guarantee finding a globally optimal prompt~\cite{reynolds2021prompt}. Similar to feature engineering or neural architecture design in ML and DL, prompt design relies on empirical evaluation, domain knowledge, and iterative refinement guided by observed model behavior rather than exhaustive search~\citep{liu2023pre, reynolds2021prompt, Prompt2023White}.  The methodology is outlined in Figure~\ref{fig:Methodology} and is described in detail in the subsequent subsections.

\subsection{Dataset Splitting}\label{sec:Dataset_Splitting}
First, a pairwise mapping is performed between each source and target artifact to create a comprehensive set of artifact pairs. Each pair was labeled as a ‘true’ or ‘false’ link based on the ground truth data, ensuring that all possible relationships are captured.

The dataset is subsequently divided into three subsets: training (Tr), validation (V), and test (T) sets. The training set provides demonstrations for few-shot learning, the validation set is utilized for prompt design, evaluation, and ranking, and the test set is reserved for the final evaluation of the top-ranked prompts. This split structure ensures clear separation for prompt design and final evaluation, preventing overfitting and providing a realistic measure of prompt effectiveness. We split the dataset with a ratio of 4:2:4 for Tr:V:T, as the training set in ICL is used only for selecting few-shot demonstrations, allowing us to maximize the test data for robust evaluation.

The dataset can be split in two ways, depending on the traceability task being simulated (i.e., TLG, TLC, or TLX):

\begin{itemize}
    \item \textbf{Split by link~\citep{guo2017semantically}:} In this approach, the dataset is split into random training, validation, and test sets, ensuring that both source and target artifacts are visible in each phase, thereby simulating the TLC task. Given the imbalance in most traceability datasets~\citep{wang2023empirical, dong2022semi}, stratified splitting should be used to maintain the same proportion of positive and negative samples in each subset. For the TLG task, the same strategy can be applied, excluding the training set, since no training data is required.

    \item \textbf{Split by artifact~\citep{lin2022enhancing}:} In this approach, the source artifacts are randomly divided into training, validation, and test sets, with all target artifacts remaining visible to each set. This approach simulates the TLX task, where new source artifacts are incrementally linked to the target artifacts.
\end{itemize}

\subsection{Core Prompt Design}
This stage focuses on formulating an effective core prompt for software artifacts traceability. The process involves designing, testing, and refining prompts iteratively based on the analysis of their output. Prompt refinement is guided by an analysis of performance on validation set, with each iteration introducing a single change evaluated quantitatively. The key steps are:

This stage focuses on formulating an effective core prompt for software artifact traceability. The process follows established prompt engineering best practices~\cite{schulhoff2024prompt,reynolds2021prompt,openai_prompt_guide}, which emphasize clear task specification, explicit definition of inputs and outputs, and iterative refinement based on the model’s observed behavior. In TraceLLM, core prompt design is treated as an iterative, human-in-the-loop process in which prompt variants are empirically evaluated on a validation set. Prompt refinement is guided by systematic analysis of the model’s outputs on the validation set, with each iteration introducing a single modification that is evaluated quantitatively. This approach is consistent with prior work showing that small changes in prompt wording or framing can have a noticeable impact on LLM behavior and performance~\cite{schulhoff2024prompt,reynolds2021prompt,openai_prompt_guide}.

The key steps of the core prompt design process are as follows:

\begin{enumerate}
    \item Designing an initial basic prompt for traceability tasks, serving as a foundation for subsequent refinements.
    \item Sampling a random source artifact and two target artifacts (one linked, one not) from the validation set to test the prompt’s ability to distinguish between true and false links.
    \item Executing the prompt by using an LLM to predict link label, then analyzing the LLM’s response to guide refinements in phrasing, structure, or focus.
    \item Repeating steps (2) and (3) for T iterations, yielding a set of candidate prompts. In this study, we set T=10 to allow for extensive refinements while accommodating the limited size of the validation set.
    \item Evaluating the candidate prompts on the full validation set and selecting the top-ranked prompt based on F2-score as the core prompt for the next stage.
\end{enumerate}

\subsection{Prompt Enrichment}

Prompt enrichment extends the core prompt by adding contextual information that guide the LLM’s interpretation of software artifacts. Based on established prompt engineering guidelines and supported by prior surveys~\cite{openai_prompt_guide, kojima2022large, wei2022chain, reynolds2021prompt, schulhoff2024prompt}, prompt enrichment in this study is explicitly structured along three dimensions: role-based context, domain context, and reasoning instructions.

Role-based enrichment assigns a specific role to the model (e.g., software traceability expert). Prior work shows that role specification can influence LLM behavior and output quality by shaping how the model interprets its task and responsibilities \cite{Prompt2023White,reynolds2021prompt,schulhoff2024prompt}.

Domain-based enrichment provides high-level domain context (e.g., aerospace systems) to help the model interpret domain-specific terms and artifacts correctly. Providing domain context is a common prompt engineering practice used to reduce ambiguity and better align the model with the task, without requiring external knowledge \cite{schulhoff2024prompt,reynolds2021prompt}.

Reasoning enrichment asks the model to explicitly reason about the relationship between artifacts before giving an answer. Prior work shows that reasoning instructions can significantly affect model behavior and performance, although their effectiveness depends on the task \cite{wei2022chain,kojima2022large,schulhoff2024prompt}.

All enriched prompt variants are evaluated on the full validation set, and the top-performing prompt is selected for the next stage.

\subsection{Evaluation}

The evaluation phase assesses the top-performing prompt on the test set to produce the final results. In addition, the prompts are evaluated on multiple LLMs and datasets to assess their generalizability and adaptability across various contexts. The results are recorded and analyzed to compare the robustness, effectiveness, and variability of the prompts, thereby identifying potential areas for refinement. 

Additionally, both zero-shot learning (for TLG) and few-shot learning (for TLC) are utilized. In zero-shot evaluation, models are evaluated on unseen data, while few-shot evaluation utilizes a limited number of training demonstrations to guide the models, providing insights into performance with limited data. However, the performance of ICL is highly sensitive to the selection and ordering of demonstrations in the prompt~\citep{peng-etal-2024-revisiting,mavromatis2023examples,lu-etal-2022-fantastically}. Selecting demonstrations randomly or without a well-defined strategy can lead to suboptimal performance, increased model variance, and inconsistent predictions~\citep{zhang-etal-2022-active,lu-etal-2022-fantastically}. 

To quantify this effect, we conducted experiments evaluating random selection of demonstrations across five different random seeds, where each demonstration set was further evaluated over five repeated runs to account for variance in LLM responses. The results revealed a noticeable variance in performance, ranging between 0.03 and 0.08. In some cases, a randomly chosen set of demonstrations significantly improved performance compared to a zero-shot experiment, demonstrating the potential benefits of demonstration selection. However, in other instances, the selected demonstrations degraded performance, making the model perform worse than zero-shot, as these demonstrations may introduce irrelevant or misleading context, effectively acting as noise. This inconsistency highlights the unreliable nature of random selection and emphasizes the need for strategies to systematically select demonstrations that improve model robustness and accuracy. In this work, we explore four major and widely used categories of demonstration selection strategies, presented in detail in Section~\ref{sec:DSSs}.

\section{Demonstrations Selection Strategies}\label{sec:DSSs}

Different demonstration selection strategies have been proposed for ICL. These strategies aim to optimize task relevance, informativeness, and generalization, ensuring that LLMs receive high-quality input for ICL. We present the most widely used demonstration selection strategies in the following sections.

\subsection{Diversity-Based Strategy}\label{sec:diversity_based_strategy}

In ICL, selecting demonstrations that are too similar to each other can limit the model’s ability to generalize across varied input patterns. Instead, diverse demonstrations with varying levels of similarity may provide a broader representation of the underlying task distribution, improving adaptability across different test queries~\citep{mavromatis2023examples}. Similarity can be lexical, structural, and semantic. While syntactic and structural diversity can be beneficial, prior research has shown that semantic diversity implicitly captures variations in linguistic structure and format~\citep{mavromatis2023examples}, making it the most effective dimension of diversity.

\subsubsection{Formulation} Given a candidate pool of labeled demonstrations \( \mathcal{D} = \{d_1, d_2, ..., d_N\} \), the goal of semantic diversity-based selection is to find a subset \( \mathcal{S} \subseteq \mathcal{D} \) of \(k\) demonstrations that maximizes semantic variance between them.

Let \( \phi(d) \) denote the embedding representation of demonstration \(d\), obtained from a pre-trained encoder (e.g., BERT). The semantic diversity objective is:

\begin{equation}
Div_{\text{semantic}}(\mathcal{S}) = \sum_{d_i \in \mathcal{S}} \sum_{d_j \in \mathcal{S}, i \neq j} (1 - \cos(\phi(d_i), \phi(d_j)))
\end{equation}

This function measures the total semantic dissimilarity of the selected demonstrations, ensuring that chosen samples are spread out in semantic space. The cosine similarity \( \cos(\phi(d_i), \phi(d_j)) \) quantifies the proximity of two demonstrations, where higher values indicate more similar meanings. By maximizing \( Div_{\text{semantic}}(\mathcal{S}) \), we encourage greater semantic contrast in the selection. Each demonstration consists of a source-target pair, which we concatenate into a single representation, for instance, \(d = \text{SOURCE\_ARTIFACT} + \text{`` ''} + \text{TARGET\_ARTIFACT}\).

The optimal subset is defined as:

\begin{equation}
\mathcal{S}^* = \arg\max_{\mathcal{S} \subseteq \mathcal{D}, |\mathcal{S}| = k} Div_{\text{semantic}}(\mathcal{S})
\end{equation}

This represents an optimization problem where the goal is to select \( k \) demonstrations that maximize semantic diversity. However, since evaluating all possible subsets is computationally infeasible, a greedy selection heuristic is employed such that new demonstrations are iteratively selected using:

\begin{equation}
d^* = \arg\min_{d \in \mathcal{D} \setminus \mathcal{S}} \sum_{d' \in \mathcal{S}} \cos(\phi(d), \phi(d'))
\end{equation}

This equation ensures that each newly selected demonstration is the least similar to those already chosen. Instead of evaluating all combinations, it efficiently picks the next demonstration that minimizes total similarity to previously selected demonstrations, thus ensuring that the set remains diverse.

\subsubsection{Implementation Algorithm} We use a greedy selection algorithm, following these steps:

\begin{algorithm}[H]
\caption{Greedy Semantic Diversity-Based Selection}
\label{alg:semantic_diversity}
\begin{algorithmic}
\State \textbf{Input:} Candidate pool $\mathcal{D}$, number of demonstrations $k$
\State \textbf{Output:} Selected diverse subset $\mathcal{S}$

\State Compute embedding $\phi(q)$ using a pre-trained encoder

\For{$d \in \mathcal{D}$}
    \State Compute embedding $\phi(d)$ using a pre-trained encoder
\EndFor

\State Compute cosine similarity matrix $M$ for all embeddings $\phi(d) \in \mathcal{D}$

\State Find the most dissimilar pair $(\phi(d_i), \phi(d_j)) = \arg\min M_{ij}$
\State $\mathcal{S} \gets \{\phi(d_i), \phi(d_j)\}$
\State $\mathcal{R} \gets \mathcal{D} \setminus \mathcal{S}$

\While{$|\mathcal{S}| < k$}
    \State Select $\phi(d^*) = \arg\min_{\phi(d) \in \mathcal{R}} \sum_{\phi(d') \in \mathcal{S}} M_{\phi(d), \phi(d')}$
    \State $\mathcal{S} \gets \mathcal{S} \cup \{\phi(d^*)\}$
    \State $\mathcal{R} \gets \mathcal{R} \setminus \{\phi(d^*)\}$
\EndWhile

\State \textbf{Return} $\mathcal{S}$
\end{algorithmic}
\end{algorithm}

The order in which demonstrations are selected and returned by the algorithm determines the sequence in which they are presented to the LLM. This ordering is preserved in the final prompt.

\subsection{Similarity-Based Selection}

In ICL, selecting demonstrations that are highly similar to the input query may improve the model’s ability to generalize to unseen data. Rather than exposing the model to a wide range of diverse cases, similarity-based selection ensures that the provided demonstrations are directly relevant to the given query~\citep{mavromatis2023examples, wang2020generalizing}. As discussed in Section~\ref{sec:diversity_based_strategy}, semantic similarity is a more effective selection criterion than lexical or syntactic similarity. Semantic similarity ensures that the selected demonstrations align with the \textit{meaning} of the query, rather than just its surface-level text, capturing both contextual and structural aspects~\citep{mavromatis2023examples}.

\subsubsection{Formulation} Given a query \( q \) and a candidate pool of labeled demonstrations \( \mathcal{D} = \{d_1, d_2, ..., d_N\} \), the goal of similarity-based selection is to find a subset \( \mathcal{S} \subseteq \mathcal{D} \) of \( k \) demonstrations that maximizes semantic alignment with \( q \).

Let \( \phi(d) \) and \( \phi(q) \) denote the embedding representations of demonstration \( d \) and query \( q \), respectively, obtained from a pre-trained encoder. The similarity objective is:

\begin{equation}
S_{\text{semantic}}(q, d) = \cos(\phi(q), \phi(d))
\end{equation} where \( \cos(\phi(q), \phi(d)) \) represents the cosine similarity between the query and demonstration embeddings. The optimal subset is defined as:

\begin{equation}
\mathcal{S}^* = \arg\max_{\mathcal{S} \subseteq \mathcal{D}, |\mathcal{S}| = k} \sum_{d \in \mathcal{S}} S_{\text{semantic}}(q, d)
\end{equation}

This ensures that the selected demonstrations are the most relevant to the query in semantic space, resulting in the optimal selection. Similarly to diversity-based selection, the source-target pair in a demonstration is concatenated into a single representation.

\subsubsection{Implementation Algorithm} We compute the similarity of each demonstration with the query and select the most relevant subset:

\begin{algorithm}[H]
\caption{Optimal Similarity-Based Selection}
\label{alg:similarity_selection}
\begin{algorithmic}
\State \textbf{Input:} Query $q$, candidate pool $\mathcal{D}$, number of demonstrations $k$
\State \textbf{Output:} Optimal similar subset $\mathcal{S}$
\For{$d \in \mathcal{D}$}
    \State Compute embedding $\phi(d)$ using a pre-trained encoder
\EndFor
\State Compute cosine similarity scores $S_{\text{semantic}}(q, d)$ for all $d \in \mathcal{D}$
\State $\mathcal{S}$ $\gets$ top $k$ most similar demonstrations
\State \textbf{Return} $\mathcal{S}$
\end{algorithmic}
\end{algorithm}

The order of demonstrations provided to the LLM is based on the similarity score to the prompt, with the most similar demonstrations appearing first.

\subsection{Uncertainty-Based Selection}

Uncertainty-based selection prioritizes demonstrations where the model shows high uncertainty, such as low-confidence predictions or frequent misclassifications, helping improve its decision boundaries~\citep{zhang-etal-2022-active}. Least confidence sampling is one of the simplest yet effective uncertainty selection methods~\citep{settles2009active}. It selects demonstrations where the model assigns the lowest probability to the correct label, forcing the model to learn from its most uncertain predictions~\citep{settles2009active}.

\subsubsection{Formulation} Given a candidate pool of labeled demonstrations \( \mathcal{D} = \{d_1, d_2, ..., d_N\} \), the goal of least confidence selection is to find a subset \( \mathcal{S} \subseteq \mathcal{D} \) of \( k \) demonstrations where the model has the lowest confidence in the true label.

Let \( y^* \) be the correct class label for demonstration \( d \), and let \( P(y^* \mid d) \) be the model’s predicted probability for that class. The confidence score is defined as:

\begin{equation}
U_{\text{LC}}(d) = P(y^* \mid d)
\end{equation} where lower \( P(y^* \mid d) \) means the model is less confident in predicting the correct label. The optimal subset is defined as:

\begin{equation}
\mathcal{S}^* = \arg\min_{\mathcal{S} \subseteq \mathcal{D}, |\mathcal{S}| = k} \sum_{d \in \mathcal{S}} U_{\text{LC}}(d)
\end{equation}

This ensures that the selected demonstrations are those where the model has the lowest confidence in the true label.

\subsubsection{Implementation Algorithm}
We use a least confidence sampling approach to select the optimal most uncertain demonstrations:

\begin{algorithm}[H]
\caption{Optimal Least Confidence-Based Selection}
\label{alg:least_confidence_selection}
\begin{algorithmic}
\State \textbf{Input:} Candidate pool $\mathcal{D}$, number of demonstrations $k$
\State \textbf{Output:} Optimal least confident subset $\mathcal{S}$
\For{$d \in \mathcal{D}$}
    \State Compute model probability $P(y^* \mid d)$ for the true label
\EndFor
\State $\mathcal{S}$ $\gets$ top $k$ least confident demonstrations
\State \textbf{Return} $\mathcal{S}$
\end{algorithmic}
\end{algorithm}

The demonstrations provided to the LLM are ordered by prediction uncertainty score, with the least certain (i.e., low-confidence) demonstrations presented first.

\subsection{Label-Aware Sampling}

Label-aware sampling ensures that the selected demonstrations maintain a balanced representation across different labels, preventing bias in model predictions. Without label balancing, the model may overfit to dominant classes, reducing its ability to generalize across underrepresented categories~\citep{du_automatic_2020}. This constraint is particularly important in classification tasks, where an unbalanced selection can result in a biased prompt, leading the model to favor frequent labels while neglecting rare ones~\citep{gao2025exploring, fei-etal-2023-mitigating}. 

This strategy is not a standalone selection strategy,  but rather a constraint that can be applied to any selection approach, including random selection. It can be seamlessly integrated into other selection strategies. For example, in a 4-shot diversity-based selection, instead of selecting the four most semantically diverse demonstrations without constraints, the algorithm would select the two most diverse positive demonstrations and the two most diverse negative demonstrations, ensuring both diversity and label balance.

\subsubsection{Formulation} Given a candidate pool of labeled demonstrations \( \mathcal{D} = \{d_1, d_2, ..., d_N\} \), where each demonstration belongs to a label \( y \in Y \), any selection strategy \( \mathcal{S} \) must satisfy:

\begin{equation}
|\mathcal{S}_y| \approx \frac{k}{|Y|}
\end{equation} for each label \( y \in Y \), ensuring that the number of selected demonstrations per label is approximately equal.

\subsubsection{Implementation Algorithm}
We enforce label balance by modifying any selection strategy to apply label-aware sampling:

\begin{algorithm}[H]
\caption{Label-Aware Sampling Constraint}
\label{alg:label_aware_constraint}
\begin{algorithmic}
\State \textbf{Input:} Candidate pool $\mathcal{D}$, selection strategy $f$, number of demonstrations $k$
\State \textbf{Output:} Label-balanced subset $\mathcal{S}$

\For{$y \in Y$}
    \State Extract $\mathcal{D}_y \subseteq \mathcal{D}$ where demonstrations have label $y$
    \State Apply selection strategy $f$ to $\mathcal{D}_y$ to obtain $\mathcal{S}_y$
    \State Select $\mathcal{S}_y = \mathcal{S}_y[: k / |Y|]$ to maintain label balance
\EndFor

\State Combine selected demonstrations: $\mathcal{S} = \bigcup\limits_{y \in Y} \mathcal{S}_y$
\State \textbf{Return} $\mathcal{S}$
\end{algorithmic}
\end{algorithm}

\subsection{Visualization of the Selected Demonstrations}

\begin{figure}[h!]
    \centering
    \renewcommand{\arraystretch}{1.2} 
    \setlength{\tabcolsep}{3pt} 
    \begin{tabular}{c c c c}
        \toprule
        & \small \textbf{Random} & \small \textbf{Diverse} & \small \textbf{Similar} \\ 
        \midrule

        \small \rotatebox{90}{\textbf{Unbalanced}} & 
        \includegraphics[width=0.3\textwidth]{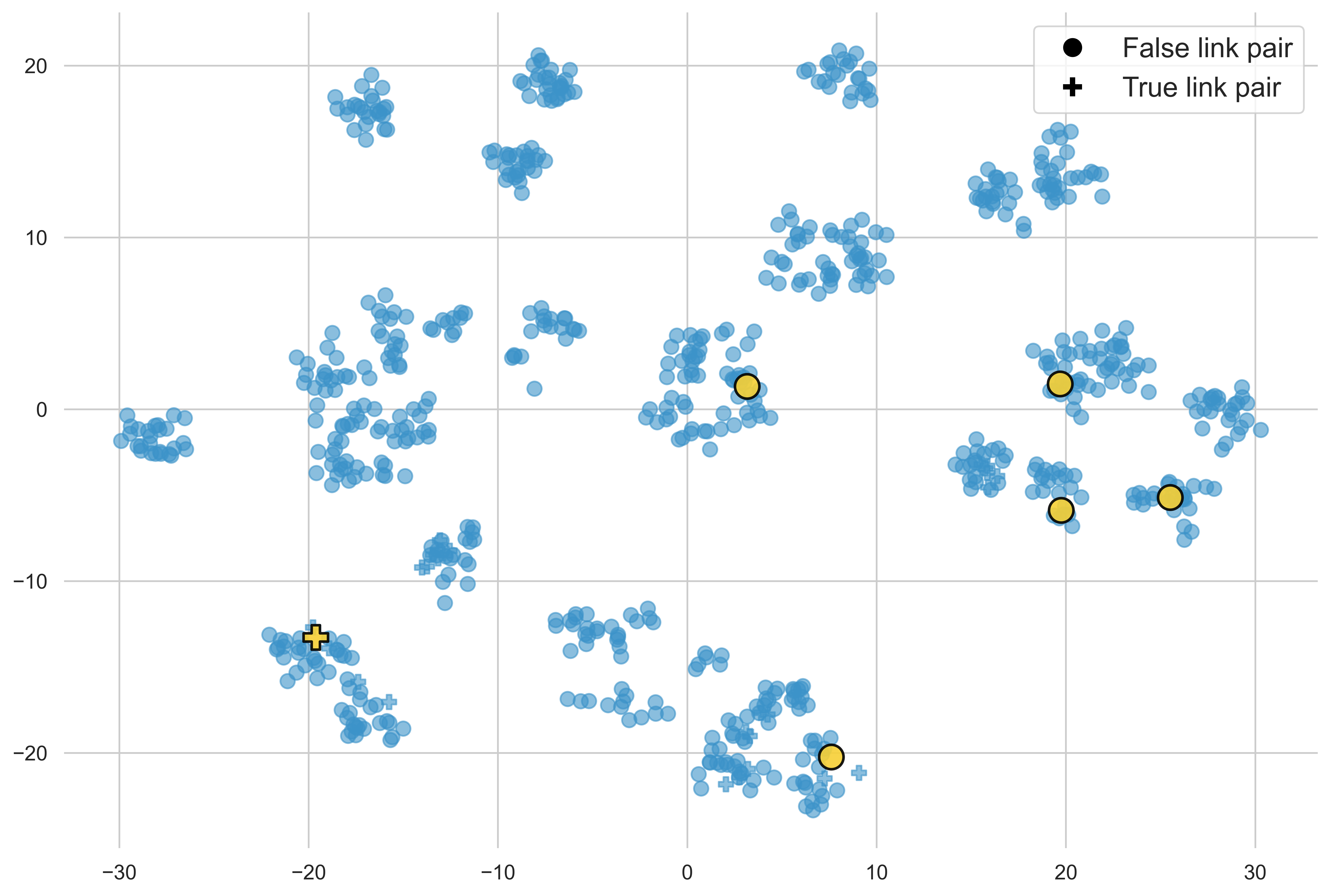} & 
        \includegraphics[width=0.3\textwidth]{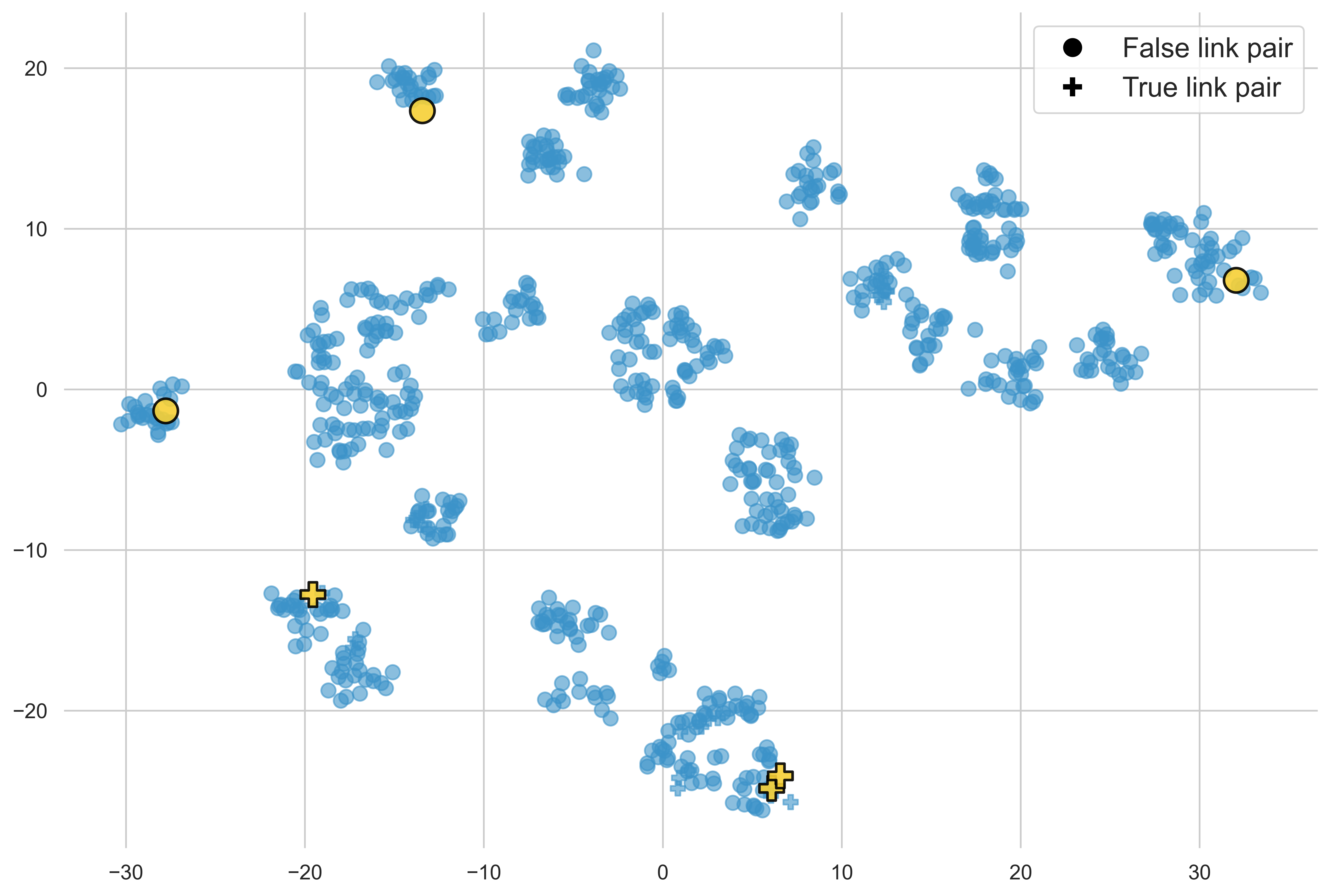} & 
        \includegraphics[width=0.3\textwidth]{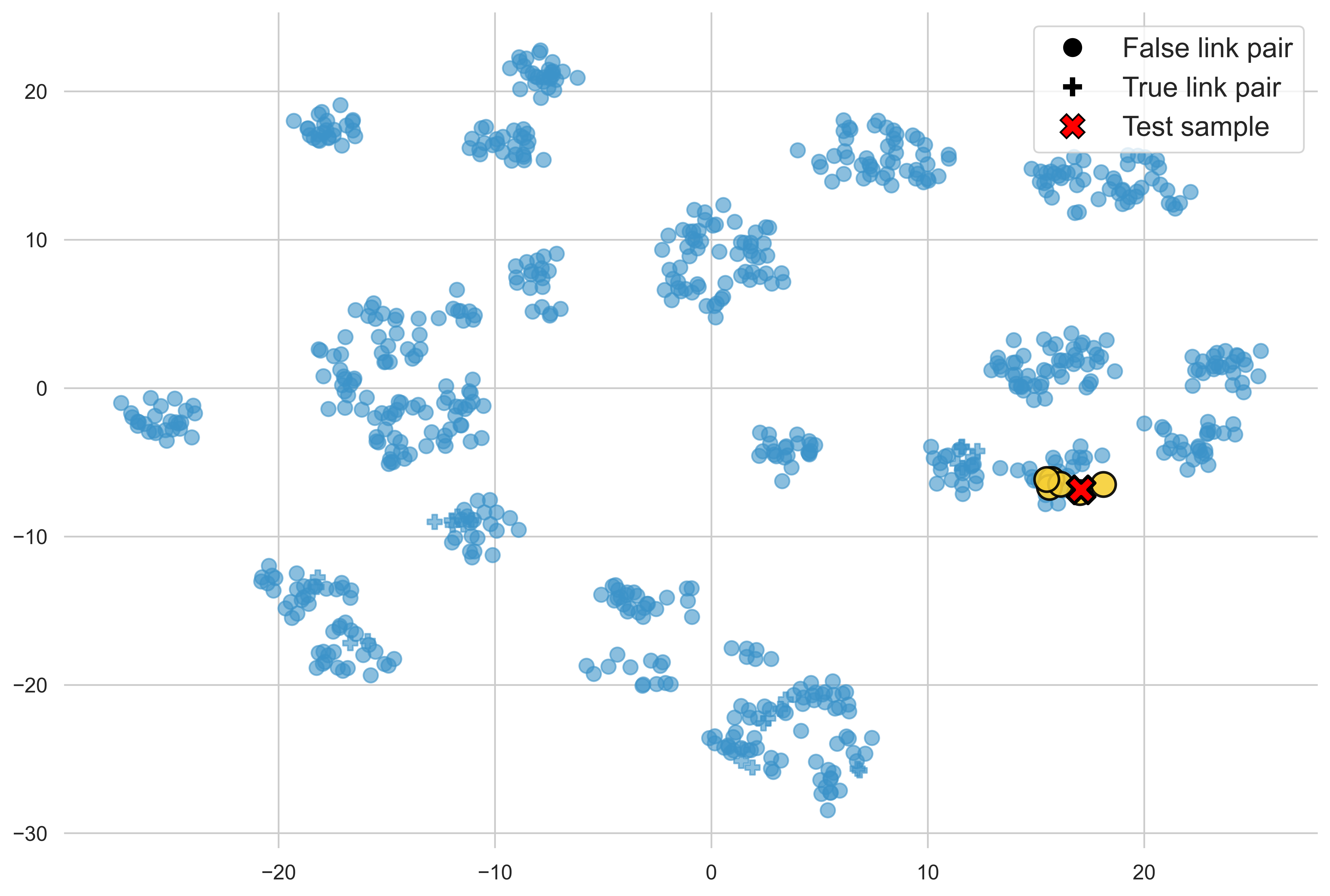} \\ 

        \midrule

        \small \rotatebox{90}{\textbf{Balanced}} & 
        \includegraphics[width=0.3\textwidth]{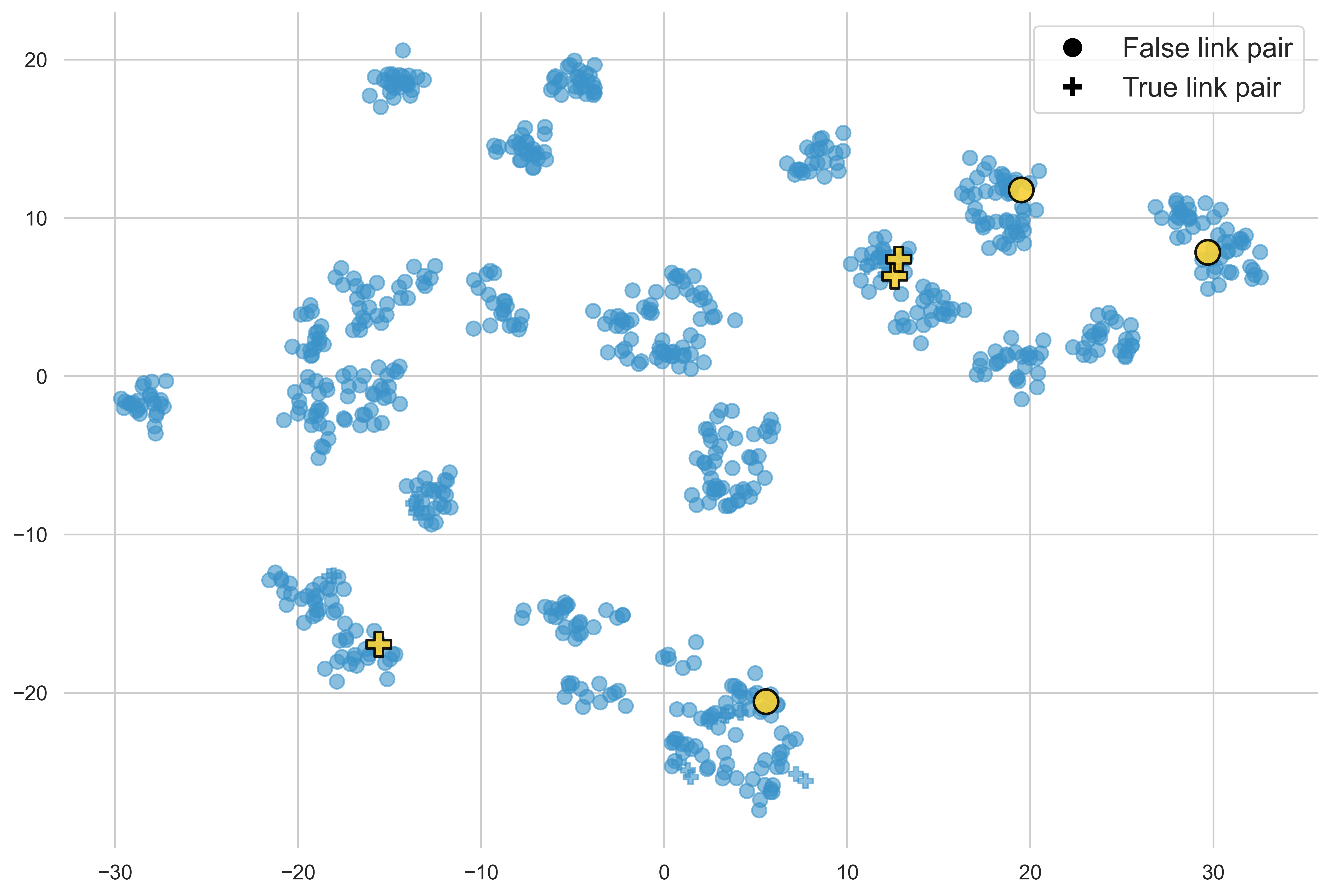} & 
        \includegraphics[width=0.3\textwidth]{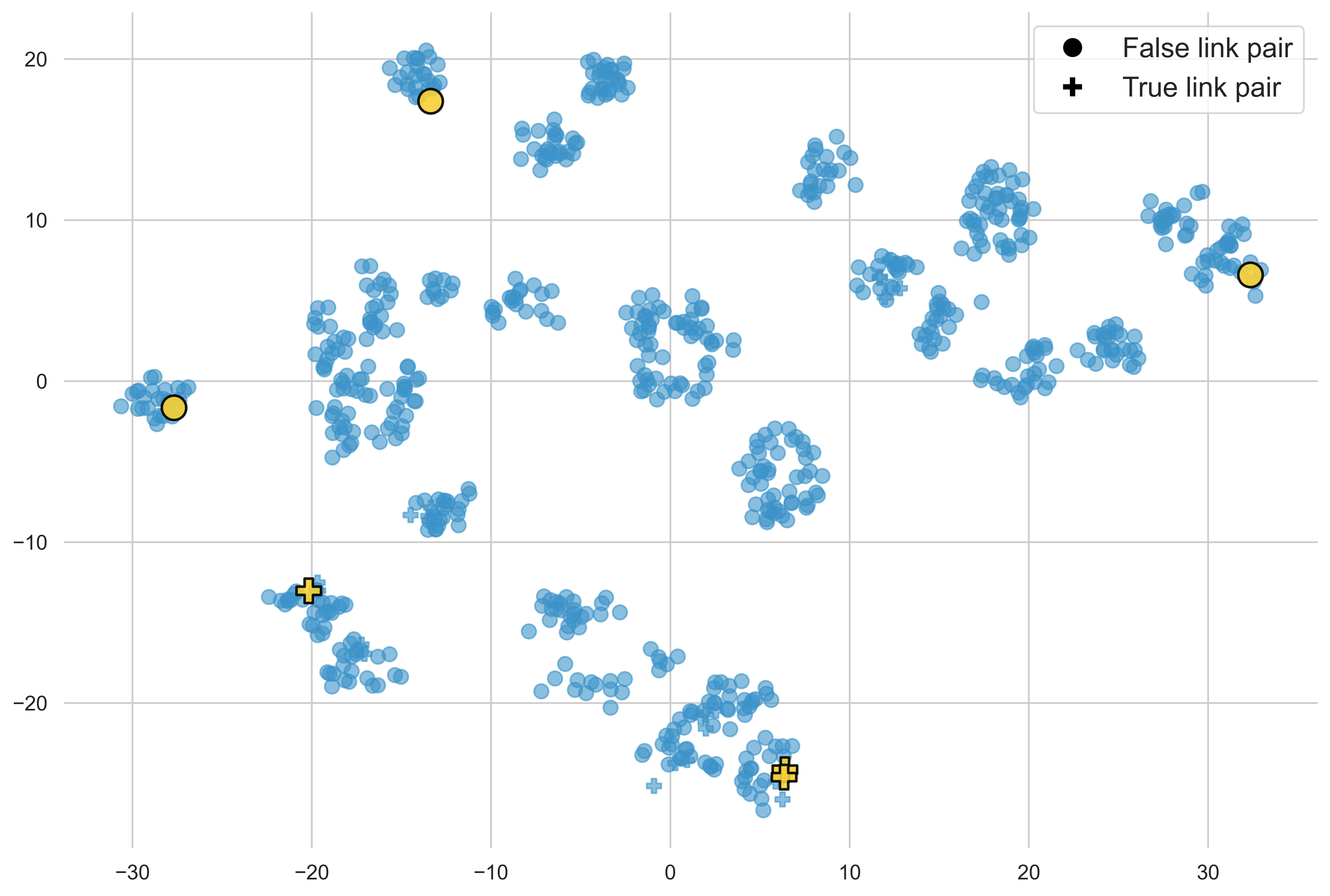} & 
        \includegraphics[width=0.3\textwidth]{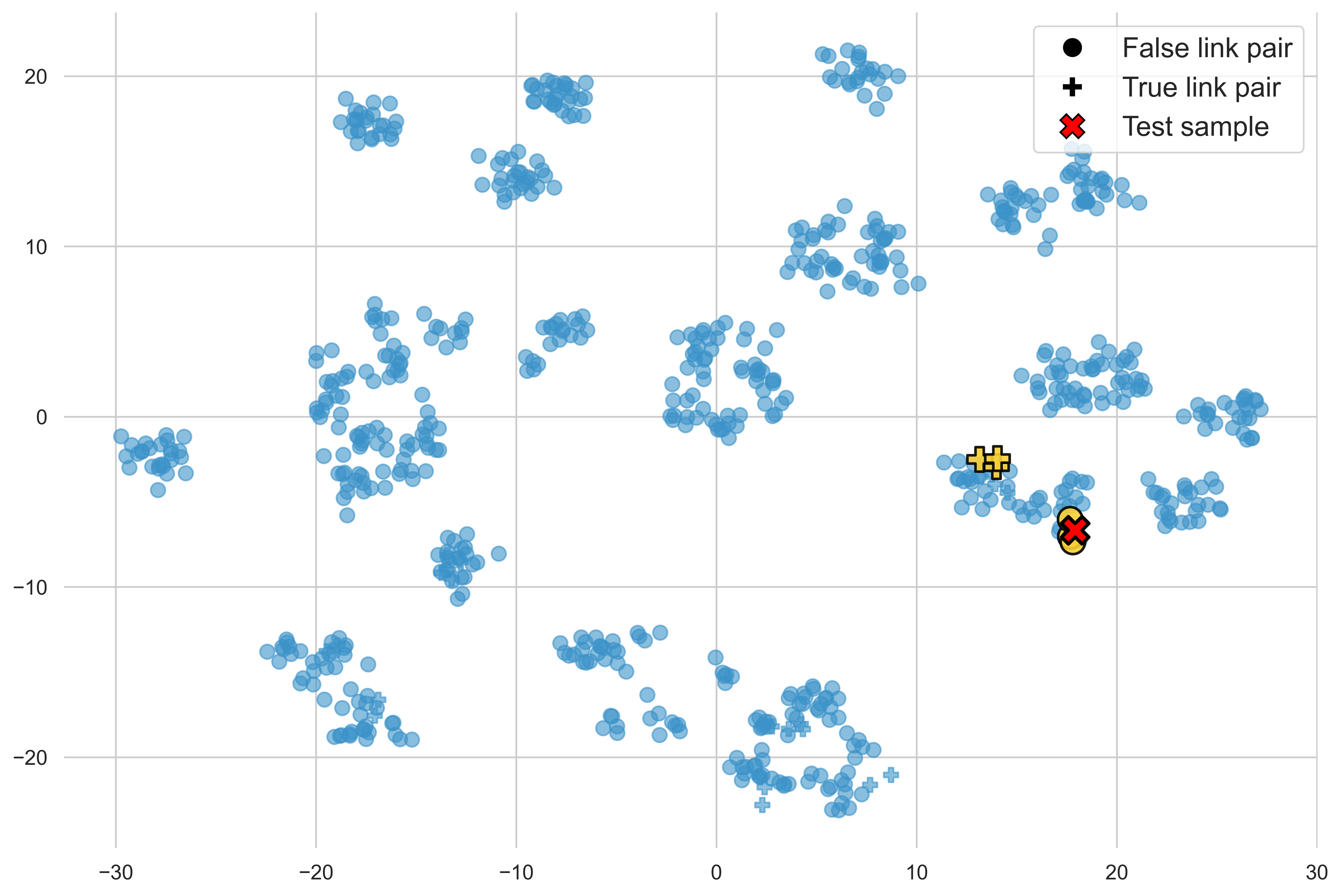} \\ 

        \bottomrule
    \end{tabular}

    \caption{T-SNE visualization of the demonstrations selected by the different selection strategies. 
    The rows indicate balanced and unbalanced selection, while the columns represent different strategies.}
    \label{fig:selection_strategies}
\end{figure}

Figure~\ref{fig:selection_strategies} presents t-SNE visualizations of the selected demonstrations based on different selection strategies. The figure is structured as a 2×3 grid, where the columns correspond to different selection strategies: Random, Diversity, and Similarity, and the rows represent the Balanced and Unbalanced label selection constraints. Each subplot illustrates the spatial distribution of selected samples within the latent space. The Uncertainty-based strategy is excluded from this visualization, as it does not rely on artifact embeddings.

In each visualization, the blue dots represent the entire training dataset, while the selected demonstrations are highlighted with yellow markers. False link pairs are denoted by circles, while true link pairs are represented by plus markers. In the Similarity strategy, a red ‘X’ marker indicates the representation of a test sample. 

A comparison between the Balanced and Unbalanced rows reveals the effect of selection balance. In the Unbalanced case, selected demonstrations tend to be false link pairs, as the dataset is highly imbalanced, where the number of false links is significantly larger than the number of true links. In contrast, in the Balanced case, the selection ensures an even distribution of labels, leading to a more representative selection of demonstrations.

The Diversity strategy results in more scattered selections compared to the random selection, covering a wider range of different demonstrations. This broader coverage may help the model generalize better by exposing it to a diverse set of patterns within the dataset. In contrast, the Random strategy results in a more arbitrary selection that may not effectively cover distinct clusters in the latent space. In addition, the selected samples in the Similarity strategy are positioned very close to the test sample in the latent space. This proximity may help the model focus on learning specific patterns relevant to the test sample, potentially leading to better adaptation.

\section{Experimental Setup}\label{sec:exp_setup}
In this section, we provide a detailed description of the experimental setup employed in this research.

\subsection{Datasets}
For this study, the CM1 dataset~\citep{hayes2006advancing} was utilized to design prompts and evaluate their effectiveness. It includes 22 high-level requirements and 53 design elements from a NASA scientific instrument, with a traceability matrix containing 45 true links out of 1,166 possible links. The requirements are typically one to two sentences, while the design elements average several paragraphs. CM1 was selected because it represents a real-world, complex software project and has been extensively used in prior traceability research~\citep{workneh2023machine, chang2023cross, rodriguez_prompts_2023, chen2019enhancing}. Its complexity arises from the use of technical domain-specific language, which offers the advantage of rigorously testing the model’s ability to handle challenging real-world scenarios.

To evaluate the generalizability of the best-performing prompt, we tested it on additional datasets: EasyClinic: Use Cases to Test Cases (UC–TC) and Use Cases to Interaction Diagrams (UC–ID) tasks~\footnote{\url{http://sarec.nd.edu/coest/datasets/EasyClinic.zip}}, as well as CCHIT dataset~\citep{cleland2010machine}. EasyClinic was selected due to its clear differences from CM1: it is less complex, originates from a different domain (healthcare), and involves distinct artifact types. CCHIT, on the other hand, maps software requirements to HIPAA (Health Insurance Portability and Accountability Act) regulations, which are mandated for all healthcare-related systems in the United States~\citep{wang2023empirical, chang2023cross, chen2019enhancing}. The CCHIT dataset is substantially larger than the others, with over 1,000 source requirements and only 10 target regulations, and poses a greater challenge due to its extreme class imbalance, with true links comprising just 1\% of all possible pairs. These datasets have been widely used in prior work~\citep{wang2023empirical, chang2023cross, chen2019enhancing}. A summary of their characteristics is provided in Table~\ref{tab:datasets}.

\begin{table}[]
\centering
\caption{Summary of datasets, including artifact types, sizes, and true link distribution (\%TL) used for evaluation.}
\label{tab:datasets}
\begin{tabular}{@{}llllllll@{}}
\toprule
\multirow{2}{*}{\textbf{Dataset}} & \multicolumn{2}{l}{\textbf{Source Artifact}} & \multicolumn{2}{l}{\textbf{Target Artifact}} & \multirow{2}{*}{\textbf{\#FL}} & \multirow{2}{*}{\textbf{\#TL}} & \multirow{2}{*}{\textbf{\%TL}} \\ \cmidrule{2-5}
 & \textbf{Type} & \textbf{Size} & \textbf{Type} & \textbf{Size} & & & \\
\midrule
CM1 & Requirement & 22 & Design element & 53 & 1121 & 45 & 4\% \\
EasyClinic(UC-TC) & Use case & 30 & Test case & 63 & 1827 & 63 & 3\% \\
EasyClinic(UC-ID) & Use case & 30 & Interaction diagram & 20 & 574 & 26 & 4\% \\
CCHIT & Requirement & 1064 & Regulation & 10 & 10562 & 78 & 1\% \\
\bottomrule
\multicolumn{8}{l}{
\footnotesize{\#FL: Number of False Links, \#TL: Number of True Links, \%TL: Percentage of True Links.}}
\end{tabular}
\end{table}

\subsection{Traceability Tasks}
In this study, we target two traceability tasks: TLG and TLC. For TLG, we use a zero-shot setting where the model generates traceability links without prior knowledge. For TLC, we adopt a few-shot learning approach with 2, 4, and 6-shot settings. In these configurations, the model is given both positive and negative trace links demonstrations to guide predictions—one of each for 2-shot, two of each for 4-shot, and three of each for 6-shot. This balanced setup helps the model to make more informed predictions. Importantly, the same demonstrations were fixed across all experiments to ensure a fair comparison. We leave the TLX task for future research.

\subsection{Large Language Models (LLMs)}

During the core prompt design and enrichment phases, we utilized GPT, a state-of-the-art large language model~\citep{hou2023large}. Specifically, GPT 4o Mini was chosen for its efficiency, making it well-suited for iterative prompt experimentation while maintaining strong performance. Once the top-ranked prompt was identified, it was evaluated on the test set using GPT 4o Mini, and then further tested across a range of recent and diverse LLMs, both lightweight and full-scale versions:including GPT 4o, Claude Haiku, Claude Sonnet, Gemini Flash, Gemini Pro, and Llama. Based on this evaluation, the best-performing model was selected, and we conducted further experiments using different demonstration selection strategies. Finally, we tested the top-performing DSS for generalizability across datasets and benchmarked it against existing baselines and state-of-the-art approaches.

For demonstration selection strategies, we used MPNet; a sentence-transformer embedding model~\citep{reimers-gurevych-2019-sentence}, to support similarity and diversity-based selection. This sentence-transformer model was chosen because of its strong performance on semantic textual similarity tasks~\citep{reimers-gurevych-2019-sentence}. It enables efficient and reliable computation of embedding distances to rank candidate demonstrations based on cosine similarity. 

\subsection{Experimental Configuration and Reporting Details}

This section summarizes the experimental configuration and reporting details required for transparency and reproducibility, following emerging guidelines for empirical software engineering studies involving LLMs~\cite{baltes2025guidelines}.

\paragraph{Model configuration.}
All experiments were conducted using the following models versions:
\textit{gpt-4o-mini}, \textit{gpt-4o}, \textit{claude-3-5-haiku-20241022}, \textit{claude-3-5-sonnet-20241022}, \textit{gemini-1.5-flash}, \textit{gemini-1.5-pro}, \textit{Llama-3.1-8B-Instruct-Turbo}, \textit{Llama-3.1-70B-Instruct-Turbo}, and \textit{all-mpnet-base-v2}. All models were accessed through OpenRouter provider~\citep{openrouter} and evaluated using identical inference parameters to ensure comparability across configurations. The temperature was set to 0.0, as commonly recommended for classification tasks~\citep{yang2024wrong}. The maximum output length to 1 token, corresponding to``Yes" or ``No". These settings were chosen to reduce output variability while maintaining consistent task behavior across models.

\paragraph{Execution window.}
All experiments were executed within the same time window (from \texttt{[February 2025]} to \texttt{[May 2025]}). This is particularly important for proprietary LLMs, whose behavior may evolve over time even when version identifiers remain unchanged.

\paragraph{Prompt usage and shot settings.}
For each experiment, we explicitly specify whether zero-shot, random few-shot, or label-aware diversity-based few-shot prompting was used. Few-shot configurations rely on a small number of demonstrations selected according to the strategies described in Section~\ref{sec:DSSs}. Prompt templates and their variants are fixed per configuration and reused consistently across models, unless otherwise stated. Full prompt templates and summaries of baseline prompts are provided in Section~\ref{sec:RQ4} and Appendix~\ref{app:baselines_configurations}.

\paragraph{Handling non-determinism.}
To account for the inherent non-determinism of LLM outputs, each prompt–artifact pair was evaluated over 25 independent runs. We report the mean and standard deviation of the resulting performance metrics across these runs. All models and baselines were evaluated under identical repetition settings to ensure a fair comparison.

\paragraph{Tool architecture.}
TraceLLM uses LLMs as standalone inference components. Prompts are sent directly to the models without additional frameworks. Aside from prompt construction and result aggregation, no external tools or retrieval mechanisms are involved in the inference process.

\subsection{Baselines}\label{sec:setup_baselines}

In this study, we employed embeddings generated from several IR-based approaches as baselines for comparison, following established methods in the literature~\citep{chang2023cross, rodriguez_prompts_2023, lin2022information}. Specifically, we used embeddings generated by VSM, LSI, and LDA. The pre-processing of text included tokenization, lowercasing, and the removal of stopwords and punctuation to ensure consistency across representations. In addition, we used embeddings generated by BERT~\citep{devlin-etal-2019-bert} model as a more advanced baseline. BERT, a transformer-based model, provides richer, context-aware embeddings that capture deeper semantic relationships between artifacts. For all baseline methods, the cosine similarity between the source and target artifact embeddings was calculated, and each pair was classified as linked or not based on a similarity threshold. Multiple threshold and hyper-parameter values were tested (as presented in Table~\ref{tab:baselines_hyperparameters}), and the threshold maximizing the F2-score was selected for final evaluation. 

\begin{table}[]
\centering
\caption{Hyper-parameters for the baseline models.}
\label{tab:baselines_hyperparameters}
\begin{tabular}{@{}llll@{}}
\toprule
\textbf{Model} & \textbf{Hyper-parameters} & \textbf{Description} & \textbf{Values} \\ \midrule
LSI & n\_components & Output dimensions & [50, 100, 150] \\ \midrule
\multirow{2}{*}{LDA} & num\_topics & Num. of latent topics & [5, 10, 20, 30] \\ 
& passes & Num. of passes & [10, 15, 20] \\ \midrule
All & similarity threshold & For classification  & 0.01-1, step 0.01 \\ \bottomrule
\end{tabular}
\end{table}

To identify relevant state-of-the-art methods, we surveyed existing traceability studies (see Section~\ref{sec:related_work}). We excluded those targeting different artifact types (e.g., code) or lacking a replication package. For studies that used similar datasets but did not provide replication or dataset partitions, we performed a qualitative comparison. This ensured that our quantitative evaluation focused on the most relevant and reproducible approaches.

First, we performed a quantitative comparison of our approach to the recently proposed ICL methods, which are:

\begin{itemize}
    \item \textbf{\citet{rodriguez_prompts_2023}}: Investigated prompt engineering for leveraging LLMs in automated traceability, using the Claude model in a zero-shot setting across requirements-to-design, requirements-to-code, and design-to-code links. However, prompts were crafted directly on the test set without a validation phase, which likely led to overestimated and unrealistic performance.
    \item \textbf{\citet{hey2025requirements}:} Proposed a RAG approach using LLMs to recover inter-requirements traceability links, evaluating two open-source and two proprietary models in a zero-shot setting for requirements-to-design and requirements-to-regulations tasks. However, the study did not include prompt engineering or investigate few-shot learning with different DSSs.
    \item \textbf{\citet{etezadi2025classification}:} Investigated an LLM-based approach guided by RICE, a prompt engineering framework, in a five-shot setting. However, the study was limited to a single LLM (GPT-4o) and focused exclusively on one artifact type (requirements to regulations). 
\end{itemize}

A detailed summary of the baseline prompt configurations, including best reported models, shot settings, and prompt templates, is provided in Appendix~\ref{app:baselines_configurations}.

As part of this study, we adopted the prompts and settings from the three recent ICL approaches and re-implemented the RAG method by \citet{hey2025requirements}. When multiple prompt variants were reported in a baseline study, we selected the variant that achieved the best performance as reported by the original authors. This ensures that all methods, including this study, were evaluated on identical training and test set partitions, allowing for a fair and consistent comparison.

Lastly, we included a qualitative comparison with state-of-the-art ML-based approaches (WELR~\citep{zhao_improved_2017}, TRAIL~\citep{mills2018automatic}, \citet{tian_adapting_2018}, and \citet{du_automatic_2020}) and DL-based methods (S2Trace~\citep{chen2019enhancing} and NLTrace~\citep{lin2022enhancing}). 

\subsection{Models Evaluation}
The evaluation of the candidate prompts across various LLMs was conducted using a set of standard metrics used in traceability studies. The used metrics are: Precision, Recall, F1 Score, and F2 Score. Our main focus will be on the F2 score, as it measures the balance between precision and recall while placing greater emphasis on recall. It is particularly valuable in traceability tasks where retrieving as many relevant links as possible is prioritized over minimizing false positives~\citep{lin2022enhancing, lin2022information, hayes2006advancing}. 

To ensure the reliability of our results, each experiment was repeated five times, and we report the average performance metrics along with their standard deviations. This approach helps account for variability in LLM responses and provides a more robust evaluation across runs. In cases where performance differences between methods are small, we increase the number of repetitions to 25 to enable meaningful statistical analysis. Statistical tests are applied in the evaluation of RQ2 and RQ3. For comparisons involving paired observations, such as evaluating the same LLM under two different prompt with similar settings, we use the Wilcoxon Signed-Rank test, a non-parametric method suitable when the assumption of normality does not hold. For comparisons across independent experimental settings (e.g., different numbers of shots), we use the Mann-Whitney U test to assess statistical significance.

\section{Results and Discussion}\label{sec:results}
This section presents, analyzes, and discusses the findings for each research question.

\subsection{RQ1: To what extent can prompt engineering improve the performance of LLMs?}

In this section, we answer RQ1 by presenting and discussing the iterative process of developing effective prompts and by analyzing the experimental results obtained from the resulting prompt. Section~\ref{sec:core_prompt_design} presents the core prompt design stage, Section~\ref{sec:prompt_enrichment} presents the prompt enrichment phase, and Section ~\ref{sec:optimized_prompt_performance} analyzes and discusses the performance of the optimized prompt. 

\subsubsection{Core Prompt Design}\label{sec:core_prompt_design}
Initially, we started with a simple prompt, but as the process evolved, we refined it to enhance the performance. Preliminary tests showed that the model frequently linked most artifacts, often resulting in an overestimation of true links. To mitigate this, we shifted the focus in this discussion towards negative samples (cases with false links) intending to enhance the model’s overall performance.

In the following, we outline the key stages of this refinement process and show how the model’s responses influenced the refinement of our prompts. We begin with the following pair of artifacts:

\begin{enumerate}[label=(\arabic*)]
    \item \textit{The DPU-CCM shall implement a mechanism whereby large memory loads and dumps can be accomplished incrementally.}
    \item \textit{Flight Software Initialization: The Command and Control CSC is initialized by spawning the CCM Control Task, ccmCtrlTask(), from the operating system startup task, usrRoot().  After the task is spawned it calls a function, ccmInit(), which creates the error/event queue...}
\end{enumerate}

\begin{figure}[h!]
    \centering
    \includegraphics[width=0.5\linewidth]{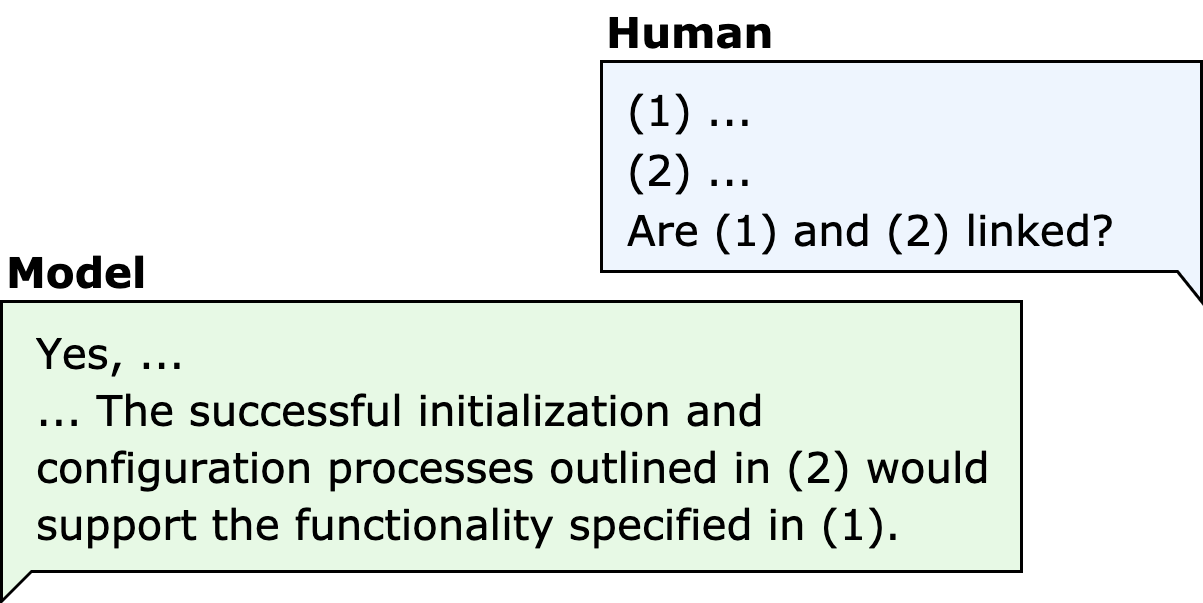}
    \caption{Initial prompt and the LLM response.}
    \label{fig:prompt_1}
\end{figure}

The initial prompt (Figure~\ref{fig:prompt_1}) was designed as a simple and direct question about whether the two artifacts are linked. However, instead of identifying the precise traceability links within the context of software engineering, the model provided general responses. For instance, it confirmed a link by stating that the initialization described in artifact (2) supported the functionality in artifact (1), suggesting that the model was interpreting ``linked'' in a broader sense. It appeared to consider any relationship as a link, rather than focusing on the software traceability context.

\begin{figure}[h!]
    \centering
    \includegraphics[width=0.5\linewidth]{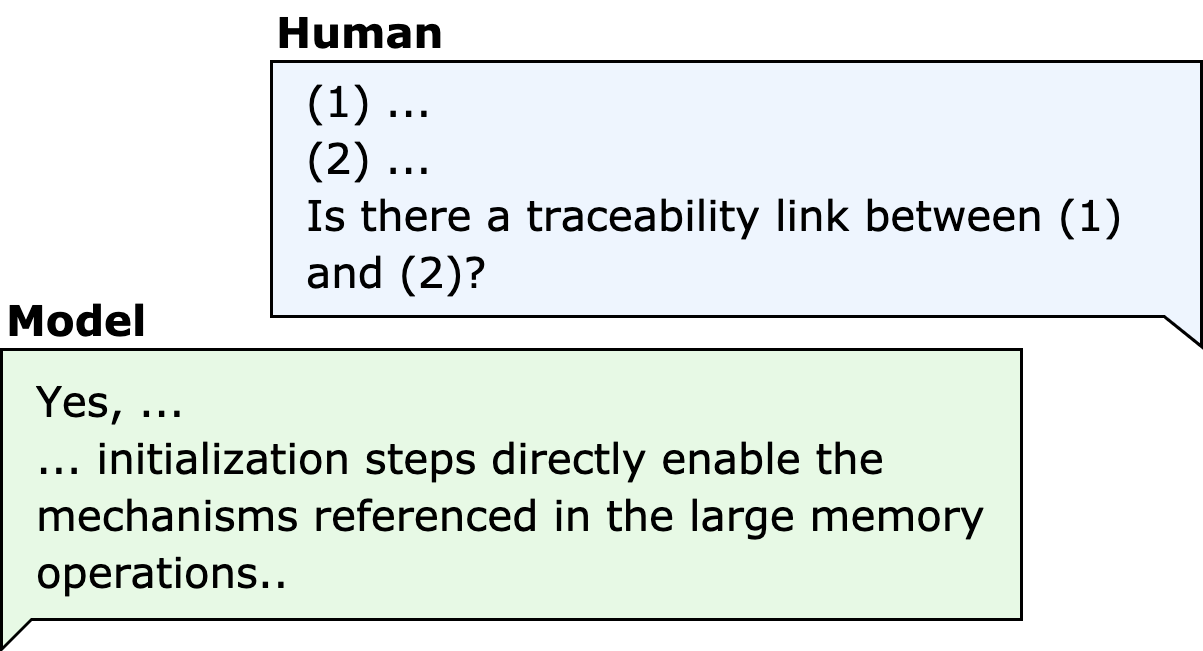}
    \caption{Prompt with context information.}
    \label{fig:prompt_2}
\end{figure}

To clarify the type of links we are interested in (those related to formal traceability in software engineering) we revised the prompt to explicitly focus on traceability links (Figure~\ref{fig:prompt_2}). By directly inquiring about ``traceability links'', we aimed to better contextualize the task. However, the model still tended to infer links where no formal traceability existed.

\begin{figure}[h!]
    \centering
    \includegraphics[width=0.5\linewidth]{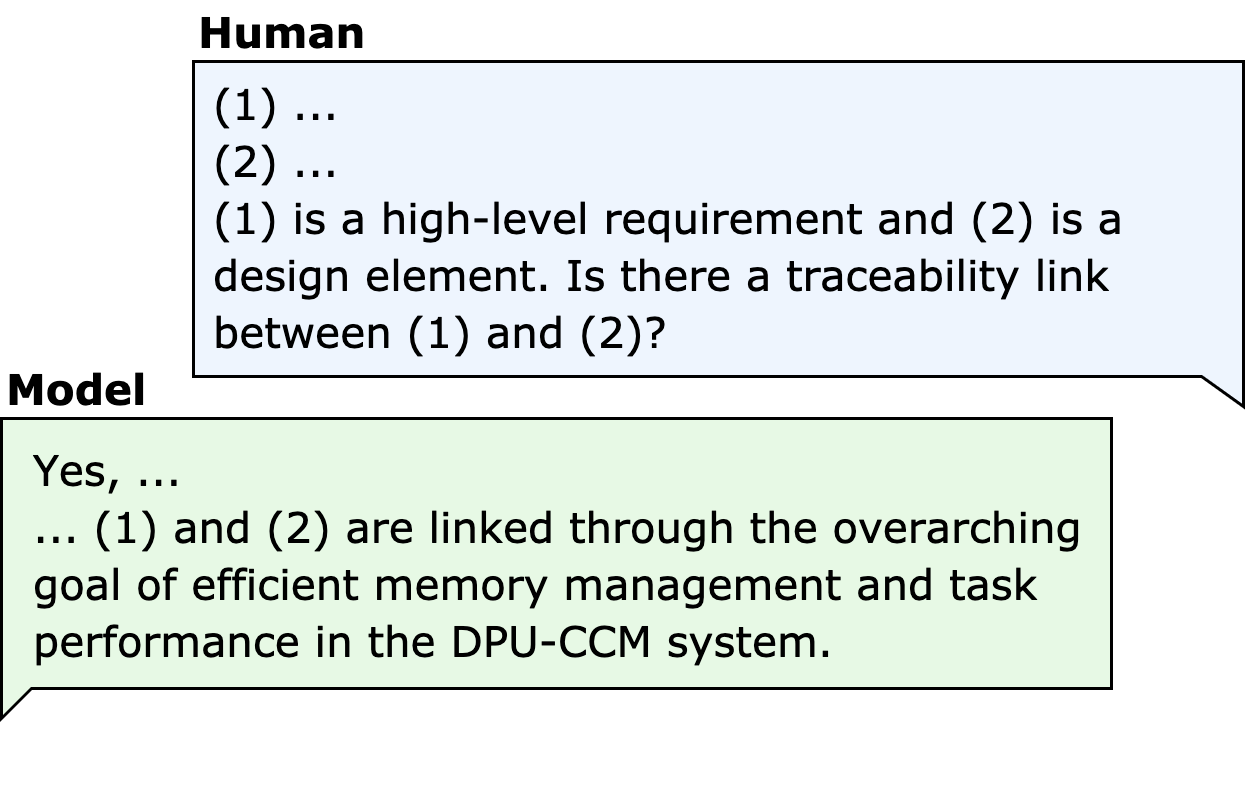}
    \caption{Prompt with artifacts-related information.}
    \label{fig:prompt_3}
\end{figure}

Subsequently, we incorporated explicit information about the nature of the artifacts into the prompt and revised it accordingly (Figure~\ref{fig:prompt_3}). By specifying that (1) is a high-level requirement and (2) is a design element, we provided the model with more precise information regarding the type of artifacts. This could allow the model to focus on the specific traceability relationships relevant to these artifact types, rather than interpreting all relationships generically. Nevertheless, challenges remained in defining how the artifacts should be connected. In particular, the model had not yet developed the ability to distinguish between different types of relationships, such as fulfillment, dependency, or testing. This highlighted the need to further enhance the model’s understanding of traceability by explicitly introducing the type of relationship between the artifacts into the prompt.

\begin{figure}[h!]
    \centering
    \includegraphics[width=0.5\linewidth]{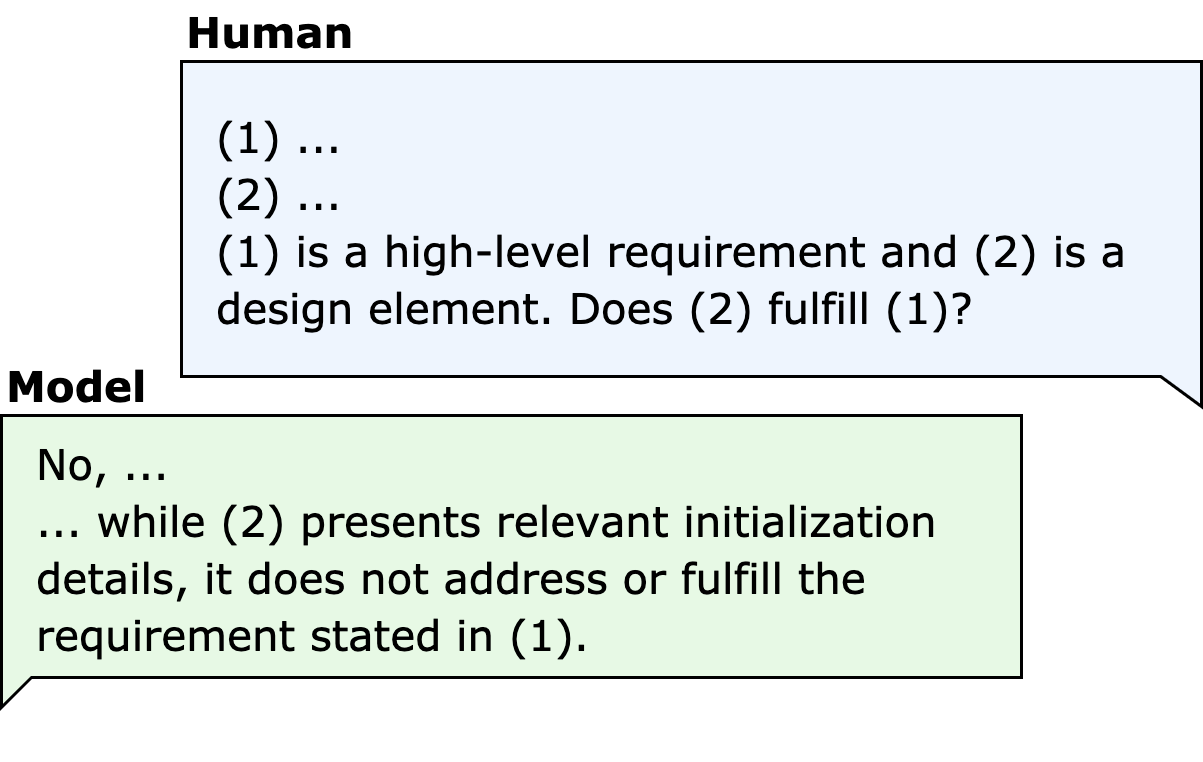}
    \caption{Prompt with relationship-related information.}
    \label{fig:prompt_4}
\end{figure}

Specifically, we instructed the model to evaluate whether the design element fulfills the high-level requirement (Figure~\ref{fig:prompt_4}). This adjustment shifted the focus toward the type of relationship, rather than its existence. By explicitly incorporating the concept of ``fulfillment'' into the prompt, we instructed the model to critically assess the relationship between the two artifacts. We experimented with various synonyms for ``fulfill'', such as ``implement'' and ``satisfy'', but ``fulfill'' produced the most accurate predictions. As a result, the model’s responses became more accurate. For instance, the model correctly identified that while artifact (2) addressed the initialization tasks, it did not fulfill the memory management requirement specified in artifact (1).

We applied the same prompt to a different pair of source and unlinked target artifacts:
\begin{enumerate}
    \item \textit{The DPU-TMALI shall be capable of making data available from the DCI to DPU-DPA.  DPU-TMALI will populate a ring buffer with frames of data for application task retrieval.}
    \item \textit{Ring buffer data structure: This definition results in the following: * administration is simple but one location is wasted in the buffer full situation, the pIn pointer is one less than the pOut pointer, * buffer empty when the pIn pointer and the pOut pointer are equal.}
\end{enumerate}

\begin{figure}[h!]
    \centering
    \includegraphics[width=0.5\linewidth]{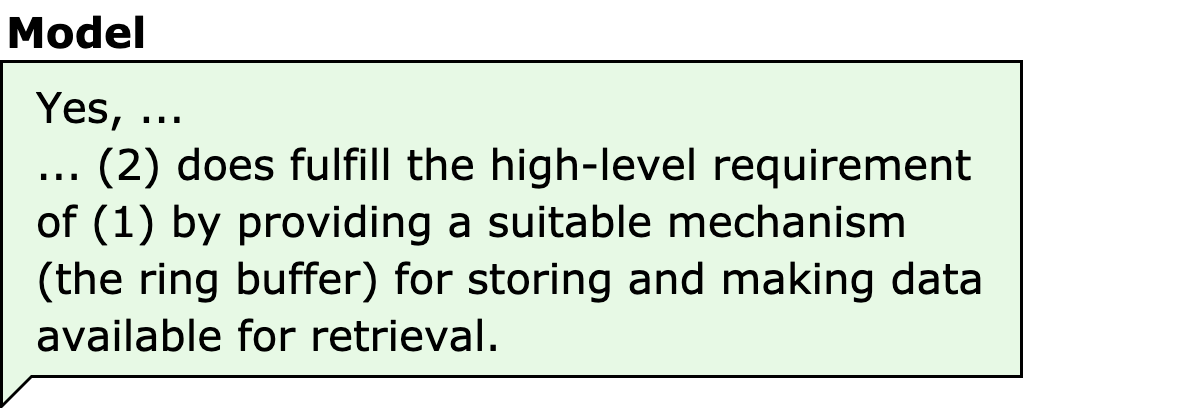}
    \caption{LLM response to unlinked artifacts.}
    \label{fig:prompt_5}
\end{figure}

In this case, however, the model incorrectly predicted a traceability link, as shown in Figure~\ref{fig:prompt_5}. The model’s response suggested that artifact (2) fulfilled artifact (1) by providing a mechanism (the ring buffer) for storing and making data available for retrieval. While both artifacts mention the ring buffer, they do not directly affect each other. This result indicated that the model was identifying indirect relationships, rather than the required direct link.

\begin{figure}[h!]
    \centering
    \includegraphics[width=0.5\linewidth]{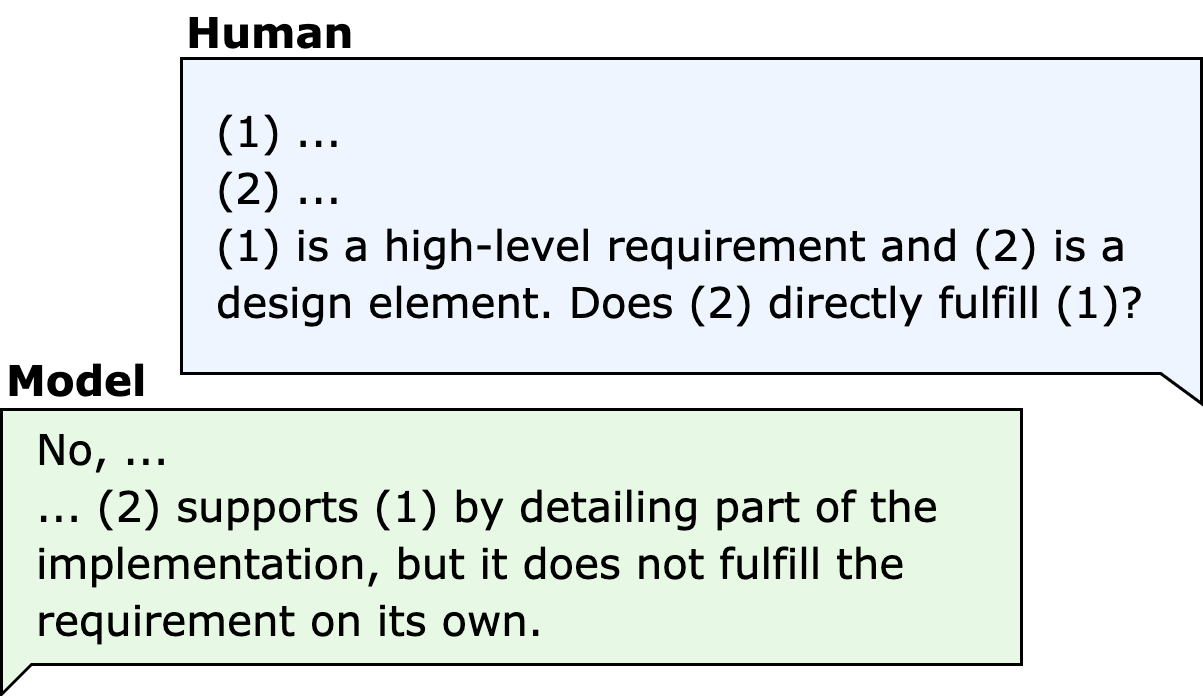}
    \caption{Prompt with enhanced relationship-related information.}
    \label{fig:prompt_6}
\end{figure}

To address this misclassification, we further refined the prompt by adding the word ``directly'' to emphasize that we were specifically seeking explicit and direct fulfillment of the requirement (Figure~\ref{fig:prompt_6}).

\begin{figure}[h!]
    \centering
    \includegraphics[width=0.7\linewidth]{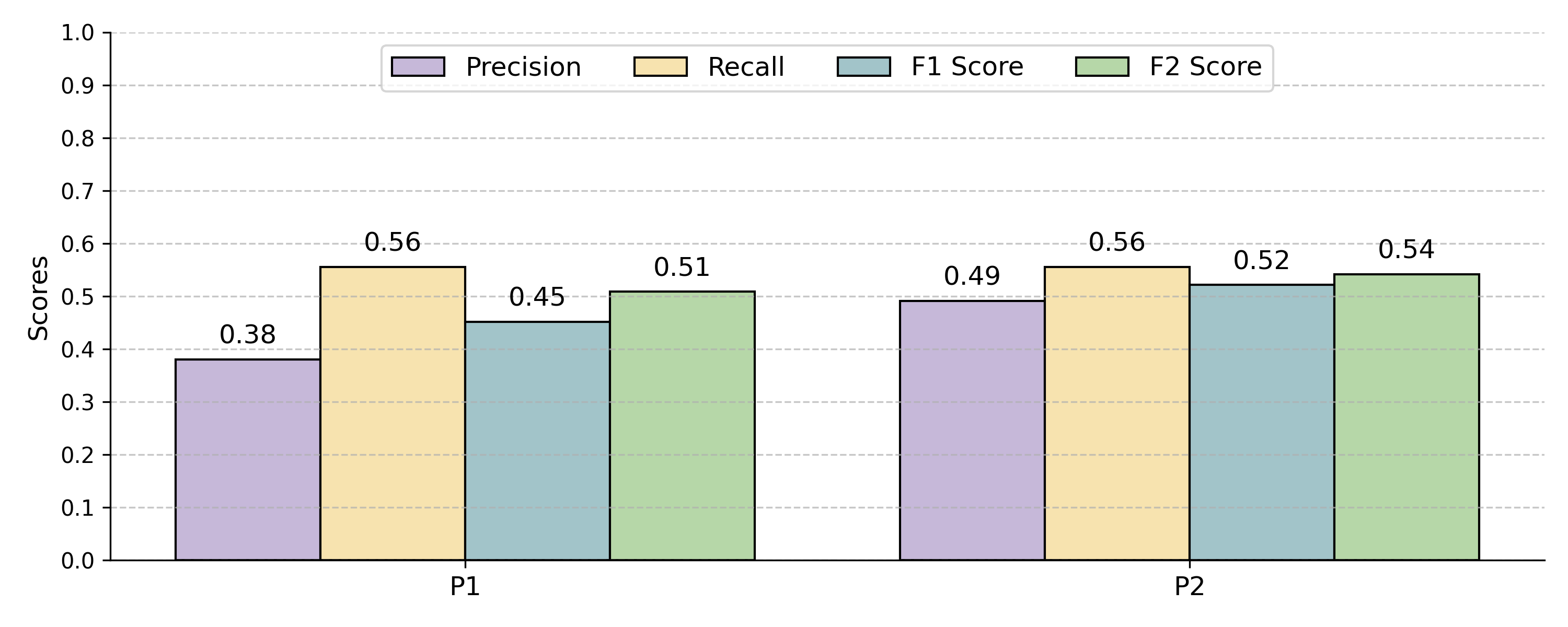}
    \caption{Performance of prompts P1 and P2.}
    \label{fig:P1_P2}
\end{figure}

Figure~\ref{fig:P1_P2} presents a bar chart comparing the performance of the two recent prompts on the validation set. The two prompts are:

\begin{enumerate}[label=\textbf{P\arabic*.}] 
    \item ``\textit{(1) is a high-level requirement and (2) is a design element. Does (2) fulfill (1)? Answer with only ‘Yes’ or ‘No’.}''
    
    \item ``\textit{(1) is a high-level requirement and (2) is a design element. Does (2) \ul{directly} fulfill (1)? Answer with only ‘Yes’ or ‘No’.}'' 
\end{enumerate}

Notably, P2 shows a significant improvement in precision, increasing from 0.38\(\pm\)0.03 in P1 to 0.49\(\pm\)0.02. This suggests that P2 applied more restrictive criteria, effectively reducing false positives. Meanwhile, recall remained consistent at 0.56\(\pm\)0.0 for both prompts, indicating that P2 did not compromise the model’s ability to detect relevant links. By maintaining recall while increasing precision, P2 achieved a more balanced performance, resulting in higher F1 (0.52\(\pm\)0.01) and F2 (0.54\(\pm\)0.01) scores.

\subsubsection{Prompt Enrichment}\label{sec:prompt_enrichment}
After identifying that the best-performing prompt is P2, we designated it as the core prompt. Then, we initiated the prompt enrichment process to enhance the model’s performance. We started by introducing specific professional roles to assess how these perspectives could influence the model’s ability to predict traceability links. The core prompt remained unchanged, however, different roles are added to the prompt to reflect diverse areas of expertise:

\begin{enumerate}[label=\textbf{P\arabic*.}, start=3] 
    \item ``\textit{\ul{You are an expert in software traceability.} (1) is a high-level requirement, and (2) is a design element. Does (2) directly fulfill (1)? Answer with only ‘Yes’ or ‘No’.}''. This represents a specific role, meant to reflect specific expertise in traceability.
    \item ``\textit{\ul{You are a software verification and validation analyst.} (1) is a high-level requirement, and (2) is a design element. Does (2) directly fulfill (1)? Answer with only ‘Yes’ or ‘No’.}''. This role was selected based on the source of the dataset~\citep{hayes2006advancing}, where verification \& validation analysts annotated the data.
    \item ``\textit{\ul{You are a requirements analyst and a software architect.} (1) is a high-level requirement, and (2) is a design element. Does (2) directly fulfill (1)? Answer with only ‘Yes’ or ‘No’.}''. This role is directly tied to the nature of the artifacts in the dataset, which include requirements and design elements.
\end{enumerate}

\begin{figure}[h!]
    \centering
    \includegraphics[width=0.7\linewidth]{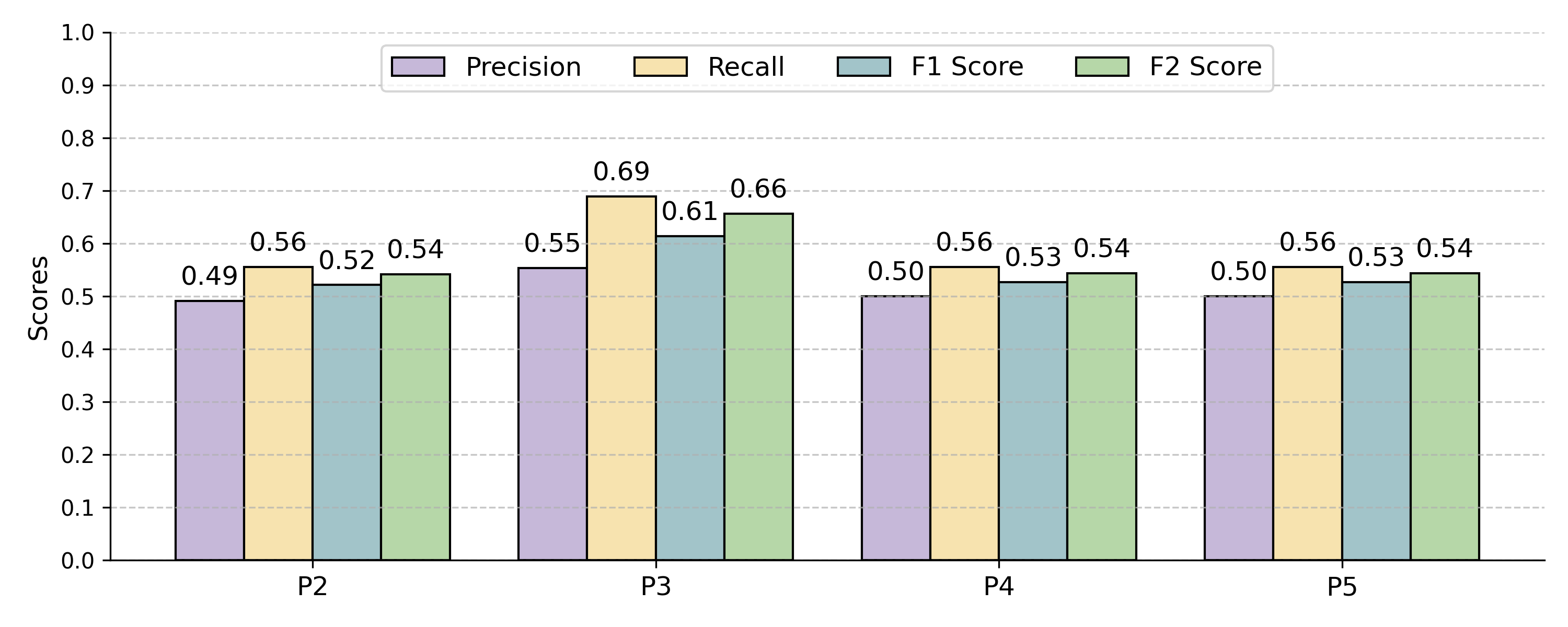}
    \caption{Performance of prompts P2, P3, P4, and P5}
    \label{fig:P2_P3_P4_P5}
\end{figure}

The bar charts in Figure~\ref{fig:P2_P3_P4_P5} present the performance metrics for the four prompts (P2–P5), highlighting how role specificity influences traceability outcomes. Among these, P3 (“You are an expert in software traceability”) achieved the highest recall (0.69\(\pm\)0.05) and precision (0.55\(\pm\)0.02), leading to the best F1 score (0.61\(\pm\)0.03) and F2 score (0.66\(\pm\)0.04). All the other three prompts: P2, P4, and P5 achieved similar performance, with a precision of \(\approx\)0.50, recall of 0.56, and F1 and F2 scores of \(\approx\)0.53 and 0.54, respectively. This indicates that a role specifically focused on traceability expertise provides the most effective balance between identifying true links and minimizing false positives.

Subsequently, we selected P3 and conducted additional experiments by incorporating contextual instructions into the prompts:

\begin{enumerate}[label=\textbf{P\arabic*.}, start=6] 
    \item ``\textit{You are an expert in software traceability. \ul{You are given two artifacts from an aerospace system.} (1) is a high-level requirement and (2) is a design element. Does (2) directly fulfill (1)? Answer with only ‘Yes’ or ‘No’.}''. This prompt incorporates domain-specific information about the artifacts, providing the model with clearer contextual grounding.
    
    \item ``\textit{You are an expert in software traceability. (1) is a high-level requirement and (2) is a design element. Does (2) directly fulfill (1)? \ul{Reason about the traceability between the two artifacts. Based on your reasoning,} answer with only ‘Yes’ or ‘No’.}''. This prompt implicitly follows a CoT approach by instructing the model to reason through the relationship before answering. The goal is to encourage deeper analysis, potentially leading to more accurate responses.
\end{enumerate}

\begin{figure}[h!]
    \centering
    \includegraphics[width=0.7\linewidth]{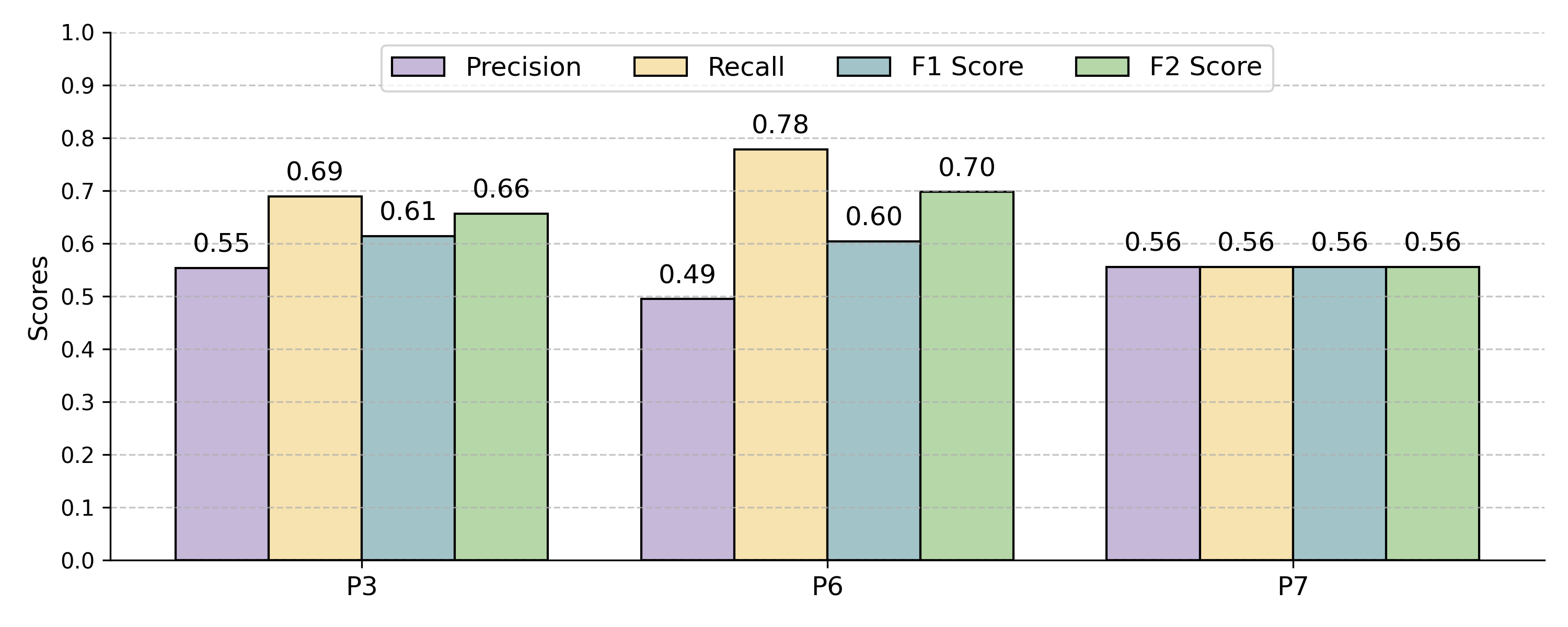}
    \caption{Performance of prompts P3, P6, and P7}
    \label{fig:P3_P6_P7}
\end{figure}

Figure~\ref{fig:P3_P6_P7} presents the results of prompts P3, P6, and P7. Prompt P6 shows a notable improvement in recall compared to P3, rising from 0.69\(\pm\)0.05 to 0.78\(\pm\)0.00, suggesting that the model captured most relevant traceability links with minimal false negatives. However, precision slightly decreased from 0.55\(\pm\)0.02 to 0.49\(\pm\)0.03, indicating that the model became less selective in identifying true traceability links. The corresponding F2-score (0.70\(\pm\)0.01) reflects the enhanced performance in terms of recall, underscoring the positive impact of incorporating domain-specific information that improves the model’s comprehension of the context and leads to more accurate predictions. On the other hand, P7 shows a substantial decrease in recall compared to P3 (from 0.69\(\pm\)0.05 to 0.56\(\pm\)0.00), with slightly high precision. However, both F1 and F2 scores decreased, leading to a poor overall performance. In summary, P6 showed the highest performance, followed by P3, with P7 showing the lowest results. We explored additional combinations of these enrichment strategies beyond those reported (e.g., combining domain context and explicit reasoning); however, these variants did not improve performance and were therefore omitted for clarity. These findings suggest that incorporating contextual information can enhance traceability link prediction. Consequently, P6 was selected as the optimized prompt.

\subsubsection{Performance Evaluation of the Optimized Prompt}\label{sec:optimized_prompt_performance}

After the prompt engineering process, we analyze the experimental results obtained from our designed prompt, P6, across different shots and random demonstrations on the test set. By focusing on F2-score, we examine how the designed prompt performs in traceability tasks. Table~\ref{tab:results_gpt_4o_mini} provides a detailed comparison of precision, recall, and F2-score of gpt-4o-mini under different shot/repeat (S/R) configurations, where S/R indicates the shot setting and run (e.g., 2/3 denotes the third run of a 2-shot setting). Metrics are reported as averages across the runs, with standard deviations to highlight performance variability. The bold value marks the highest F2-score.

\begin{table}[]
\centering
\caption{Performance of the prompt P6 using GPT-4o-Mini across zero and few-shots settings on CM1 dataset.}
\label{tab:results_gpt_4o_mini}
\begin{tabular}{@{}llll@{}}
\toprule

\textbf{S/R} & \textbf{Precision} & \textbf{Recall} & \textbf{F2-score} \\ \midrule
0/1 & 0.52\(\pm\)0.03 & 0.59\(\pm\)0.03 & 0.57\(\pm\)0.03 \\ \midrule

2/1 & 0.43\(\pm\)0.02 & 0.67\(\pm\)0.03 & 0.60\(\pm\)0.01 \\
2/2 & 0.43\(\pm\)0.02 & 0.60\(\pm\)0.02 & 0.56\(\pm\)0.02 \\
2/3 & 0.36\(\pm\)0.02 & 0.56\(\pm\)0.00 & 0.50\(\pm\)0.01 \\
2/4 & 0.48\(\pm\)0.01 & 0.50\(\pm\)0.00 & 0.50\(\pm\)0.00 \\
2/5 & 0.39\(\pm\)0.02 & 0.56\(\pm\)0.04 & 0.51\(\pm\)0.03 \\
\cmidrule{2-4}
Avg. & 0.42\(\pm\)0.04 & 0.58\(\pm\)0.06 & 0.53\(\pm\)0.04 \\
\midrule

4/1 & 0.41\(\pm\)0.02 & 0.67\(\pm\)0.00 & 0.59\(\pm\)0.01 \\
4/2 & 0.51\(\pm\)0.01 & 0.56\(\pm\)0.00 & 0.54\(\pm\)0.00 \\
4/3 & 0.45\(\pm\)0.00 & 0.50\(\pm\)0.00 & 0.49\(\pm\)0.00 \\
4/4 & 0.41\(\pm\)0.01 & 0.69\(\pm\)0.03 & 0.60\(\pm\)0.02 \\
4/5 & 0.55\(\pm\)0.03 & 0.50\(\pm\)0.00 & 0.51\(\pm\)0.01 \\
\cmidrule{2-4}
Avg. & 0.46\(\pm\)0.06 & 0.58\(\pm\)0.08 & 0.55\(\pm\)0.05 \\
\midrule

6/1 & 0.45\(\pm\)0.02 & 0.64\(\pm\)0.03 & 0.59\(\pm\)0.03 \\
6/2 & 0.53\(\pm\)0.01 & 0.61\(\pm\)0.00 & 0.59\(\pm\)0.00 \\
6/3 & 0.37\(\pm\)0.03 & 0.58\(\pm\)0.03 & 0.52\(\pm\)0.03 \\
6/4 & 0.45\(\pm\)0.01 & 0.67\(\pm\)0.00 & 0.61\(\pm\)0.01 \\
6/5 & 0.51\(\pm\)0.03 & 0.62\(\pm\)0.02 & 0.59\(\pm\)0.02 \\
\cmidrule{2-4}
Avg. & 0.46\(\pm\)0.05 & 0.62\(\pm\)0.03 & \textbf{0.58\(\pm\)0.03} \\
\bottomrule
\end{tabular}
\end{table}

In the 0-shot setting, the model achieved 0.57 F2-score. In the few-shot experiments, we observe that the choice of demonstrations highly affects the model's performance (with standard deviation ranging from 0.04 to 0.08). In the 2-shot setting, the average F2-score is lower than the 0-shot model (0.53 vs. 0.57). However, some 2-shot runs achieved higher performance (such as 2/1 with F2-score of 0.60). This suggests that the selection of demonstrations is crucial, which is explored in the answer of RQ3. The highest average performance is achieved by 3-shot setting, with F2-score of the 0.58.  

\vspace{5pt}
\noindent\fbox{
    \parbox{0.95\columnwidth}{
        \textbf{Observation 1:} Few-shot learning generally enhances performance, although its success depends heavily on the selection of demonstrations.}
}
\vspace{5pt}

\subsection{RQ2: To what extent does a prompt designed for one LLM generalize to other LLMs, compared to LLM-specific prompts?}

This section answers RQ2 by analyzing the generalization of P6 across various LLMs, shot settings, and random demonstrations in few-shot learning. Additionally, we applied TraceLLM on each model individually, designing custom prompts optimized for their specific behavior (see Table~\ref{tab:llm_specific_prompts}). We then compared each model’s performance using the re-used P6 prompt against its performance with the LLM-specific prompt.

\begin{table}[]
\centering
\small
\caption{LLM-specific prompts used in this study.}
\label{tab:llm_specific_prompts}
\begin{tabular}{@{}ll@{}}
\toprule
\textbf{Model} & \textbf{Prompt} \\
\midrule
GPT-4o-Mini & \begin{tabular}[c]{@{}l@{}}
You are an expert in software traceability. You are given two artifacts\\
from an aerospace system. (1) is a high-level requirement and (2) is a\\
design element. Does (2) directly fulfill (1)? Answer with only `Yes' or\\
`No'.\end{tabular} \\ \midrule

GPT-4o & \begin{tabular}[c]{@{}l@{}}
You are an expert in software traceability. (1) is a high-level\\ requirement and (2) is a design element. Does (2) directly fulfill (1)?\\ Reason about the traceability between the two artifacts. Based\\ on your reasoning, answer with only `Yes' or `No'. \end{tabular} \\ \midrule

Claude 3.5 Haiku & \begin{tabular}[c]{@{}l@{}}
(1) is a high-level requirement and (2) is a design element. Does (2)\\
directly fulfill (1)? Answer with only `Yes' or `No'. \end{tabular} \\\midrule

Claude 3.5 Sonnet & \begin{tabular}[c]{@{}l@{}}
(1) is a high-level requirement and (2) is a design element. Does (2)\\ fulfill (1)? Answer with only `Yes' or `No'. \end{tabular}\\\midrule

Gemini 1.5 Flash & \begin{tabular}[c]{@{}l@{}}
You are given two artifacts from an aerospace system. (1) is a high-\\level requirement and (2) is a design element. Does (2) fulfill (1) \\ Answer with only `Yes' or `No'.\end{tabular} \\ \midrule

Gemini 1.5 Pro & \begin{tabular}[c]{@{}l@{}}
(1) is a high-level requirement and (2) is a design element. Does (2)\\
directly fulfill (1)? Answer with only `Yes' or `No'. \end{tabular}\\\midrule

LLaMA 3.1 8B & \begin{tabular}[c]{@{}l@{}}
You are an expert in software traceability. (1) is a high-level\\
requirement and (2) is a design element. Does (2) fulfill (1)?\\
Answer with only `Yes' or `No'.\end{tabular} \\\midrule

LLaMA 3.1 70B & \begin{tabular}[c]{@{}l@{}}
You are an expert in software traceability. (1) is a high-level\\
requirement and (2) is a design element. Does (2) fulfill (1)?\\
Answer with only `Yes' or `No'. \end{tabular}\\
\bottomrule
\end{tabular}
\end{table}

The results are shown in Table~\ref{tab:reused_specific_comparison}. Each sub-table reports the performance of a single LLM across two groups, one using the reused prompt and the other using an LLM-specific prompt, evaluated under different shot settings. The last row of each sub-table presents the p-value indicating the statistical significance of the difference between the best F2-scores from each group, using a significance level of 0.05 (i.e., 95\% confidence). In cases where the F2-scores were equal, the configuration with fewer demonstrations is preferred.

\begin{table*}[h!]
\centering
\caption{Performance results of reused and LLM-specific prompts across different models.}
\label{tab:reused_specific_comparison}
\begin{adjustbox}{max width=\textwidth}
\renewcommand{\arraystretch}{1.3}
\begin{tabular}{@{}c|lll lll|lll lll@{}}

\Xhline{0.8pt}
\multirow{2}{*}{\textbf{Shot}} &
\multicolumn{3}{l}{\textbf{Re-used Prompt}} &
\multicolumn{3}{l|}{\textbf{LLM-Specific Prompt}} &
\multicolumn{3}{l}{\textbf{Re-used Prompt}} &
\multicolumn{3}{l}{\textbf{LLM-Specific Prompt}}\\
\cmidrule{2-13}
& \textbf{Precision} & \textbf{Recall} & \textbf{F2-Score} & \textbf{Precision} & \textbf{Recall} & \textbf{F2-Score} &
\textbf{Precision} & \textbf{Recall} & \textbf{F2-Score} & \textbf{Precision} & \textbf{Recall} & \textbf{F2-Score}\\
\hline


\rowcolor{gray!15} & \multicolumn{6}{c|}{\textbf{GPT-4o}} & \multicolumn{6}{c}{\textbf{Claude 3.5 Haiku}} \\ \hline
0 & 
0.57\(\pm\)0.05 & 0.44\(\pm\)0.06 & 0.46\(\pm\)0.06 & 
0.60\(\pm\)0.03 & 0.44\(\pm\)0.04 & 0.46\(\pm\)0.04 & 

0.15\(\pm\)0.00 & 0.78\(\pm\)0.00 & 0.42\(\pm\)0.00 & 
0.35\(\pm\)0.00 & 0.61\(\pm\)0.00 & \underline{\textbf{0.53\(\pm\)0.00}} \\


2 & 
0.47\(\pm\)0.07 & 0.54\(\pm\)0.02 & 0.52\(\pm\)0.03 &
0.46\(\pm\)0.02 & 0.55\(\pm\)0.01 & 0.53\(\pm\)0.01 & 

0.17\(\pm\)0.06 & 0.86\(\pm\)0.05 & \textbf{0.46\(\pm\)0.08} &
0.24\(\pm\)0.09 & 0.76\(\pm\)0.08 & 0.50\(\pm\)0.08 \\


4 & 
0.45\(\pm\)0.04 & 0.56\(\pm\)0.03 & 0.53\(\pm\)0.03 &
0.44\(\pm\)0.02 & 0.56\(\pm\)0.01 & 0.54\(\pm\)0.01 & 

0.17\(\pm\)0.04 & 0.84\(\pm\)0.04 & 0.46\(\pm\)0.06 &
0.21\(\pm\)0.07 & 0.76\(\pm\)0.06 & 0.48\(\pm\)0.09 \\


6 & 
0.44\(\pm\)0.03 & 0.57\(\pm\)0.01 & \textbf{0.54\(\pm\)0.02} &
0.48\(\pm\)0.02 & 0.61\(\pm\)0.01 & \underline{\textbf{0.58\(\pm\)0.01}} & 

0.14\(\pm\)0.04 & 0.79\(\pm\)0.05 & 0.41\(\pm\)0.06 &
0.17\(\pm\)0.03 & 0.74\(\pm\)0.03 & 0.45\(\pm\)0.04 \\ 
\hline

\multicolumn{2}{l}{\textbf{Best F2-Scores}} 
& & Difference & +0.04 & P-Value & \cellcolor{green}0.0017 & 
& & Difference & +0.07 & P-Value & \cellcolor{green}0.0001\\

\hline


\rowcolor{gray!10} & \multicolumn{6}{c|}{\textbf{Claude 3.5 Sonnet}} & \multicolumn{6}{c}{\textbf{Gemini 1.5 Flash}} \\ \hline
0 & 
0.34\(\pm\)0.00 & 0.72\(\pm\)0.00 & \underline{\textbf{0.58\(\pm\)0.00}} & 
0.39\(\pm\)0.00 & 0.61\(\pm\)0.00 & 0.55\(\pm\)0.00 & 

0.50\(\pm\)0.00 & 0.50\(\pm\)0.00 & 0.50\(\pm\)0.00 & 
0.46\(\pm\)0.00 & 0.67\(\pm\)0.00 & \underline{\textbf{0.61\(\pm\)0.00}} \\


2 & 
0.34\(\pm\)0.09 & 0.64\(\pm\)0.03 & 0.54\(\pm\)0.03 &
0.37\(\pm\)0.12 & 0.62\(\pm\)0.05 & 0.53\(\pm\)0.06 & 

0.42\(\pm\)0.08 & 0.53\(\pm\)0.04 & 0.50\(\pm\)0.04 &
0.39\(\pm\)0.04 & 0.67\(\pm\)0.07 & 0.58\(\pm\)0.03 \\


4 & 
0.39\(\pm\)0.06 & 0.66\(\pm\)0.04 & 0.58\(\pm\)0.03 &
0.45\(\pm\)0.07 & 0.62\(\pm\)0.06 & \textbf{0.57\(\pm\)0.03} & 

0.55\(\pm\)0.07 & 0.57\(\pm\)0.04 & \textbf{ 0.56\(\pm\)0.03} &
0.41\(\pm\)0.04 & 0.64\(\pm\)0.06 & 0.58\(\pm\)0.05 \\


6 & 
0.40\(\pm\)0.09 & 0.64\(\pm\)0.03 & 0.57\(\pm\)0.04 &
0.40\(\pm\)0.07 & 0.61\(\pm\)0.00 & 0.55\(\pm\)0.03 & 

0.62\(\pm\)0.10 & 0.51\(\pm\)0.08 & 0.52\(\pm\)0.06 &
0.43\(\pm\)0.06 & 0.60\(\pm\)0.05 & 0.55\(\pm\)0.03 \\ 
\hline

\multicolumn{2}{l}{\textbf{Best F2-Scores}} 
& & Difference & -0.01 & P-Value & \cellcolor{red}0.1972 & 
& & Difference & +0.05 & P-Value & \cellcolor{green}0.0000\\

\hline


\rowcolor{gray!10} & \multicolumn{6}{c|}{\textbf{Gemini 1.5 Pro}} & \multicolumn{6}{c}{\textbf{LLaMA 3.1 8B}} \\ \hline
0 & 
0.37\(\pm\)0.00 & 0.39\(\pm\)0.00 & 0.38\(\pm\)0.00 & 
0.35\(\pm\)0.00 & 0.44\(\pm\)0.00 & 0.42\(\pm\)0.00 & 

1.00\(\pm\)0.00 & 0.06\(\pm\)0.00 & 0.07\(\pm\)0.00 & 
0.43\(\pm\)0.01 & 0.39\(\pm\)0.00 & 0.40\(\pm\)0.00 \\


2 & 
0.36\(\pm\)0.06 & 0.69\(\pm\)0.05 & \textbf{0.57\(\pm\)0.04} &
0.38\(\pm\)0.08 & 0.68\(\pm\)0.04 & \underline{\textbf{0.58\(\pm\)0.04}} & 

0.22\(\pm\)0.07 & 0.64\(\pm\)0.18 & \underline{\textbf{0.45\(\pm\)0.09}} &
0.29\(\pm\)0.08 & 0.56\(\pm\)0.09 & \underline{\textbf{0.45\(\pm\)0.05}} \\


4 & 
0.33\(\pm\)0.03 & 0.71\(\pm\)0.04 & 0.57\(\pm\)0.01 &
0.30\(\pm\)0.03 & 0.66\(\pm\)0.07 & 0.53\(\pm\)0.04 & 

0.19\(\pm\)0.05 & 0.71\(\pm\)0.09 & 0.44\(\pm\)0.07 &
0.22\(\pm\)0.06 & 0.63\(\pm\)0.12 & 0.44\(\pm\)0.04 \\


6 & 
0.27\(\pm\)0.05 & 0.75\(\pm\)0.08 & 0.54\(\pm\)0.04 &
0.26\(\pm\)0.05 & 0.69\(\pm\)0.06 & 0.51\(\pm\)0.04 & 

0.10\(\pm\)0.04 & 0.74\(\pm\)0.11 & 0.30\(\pm\)0.06 &
0.13\(\pm\)0.07 & 0.69\(\pm\)0.11 & 0.34\(\pm\)0.06 \\ 
\hline

\multicolumn{2}{l}{\textbf{Best F2-Scores}} 
& & Difference & +0.01 & P-Value & \cellcolor{red}0.6528 & 
& & Difference & 0.00 & P-Value & \cellcolor{red}0.5782\\

\hline


\rowcolor{gray!10} & \multicolumn{6}{c|}{\textbf{LLaMA 3.1 70B}} & \multicolumn{6}{c}{} \\ \cmidrule{1-7}
0 & 
0.46\(\pm\)0.00 & 0.33\(\pm\)0.00 & 0.35\(\pm\)0.00 & 
0.36\(\pm\)0.01 & 0.44\(\pm\)0.00 & 0.42\(\pm\)0.00 & 

& & & & & \\


2 & 
0.34\(\pm\)0.03 & 0.60\(\pm\)0.01 & 0.52\(\pm\)0.01 &
0.32\(\pm\)0.03 & 0.53\(\pm\)0.07 & 0.47\(\pm\)0.05 &

& & & & & \\


4 & 
0.39\(\pm\)0.06 & 0.66\(\pm\)0.04 & \underline{\textbf{0.58\(\pm\)0.03}} &
0.45\(\pm\)0.07 & 0.62\(\pm\)0.06 & \textbf{0.57\(\pm\)0.03} & 

& & & & &\\


6 & 
0.36\(\pm\)0.05 & 0.60\(\pm\)0.01 & 0.53\(\pm\)0.02 &
0.34\(\pm\)0.04 & 0.60\(\pm\)0.03 & 0.52\(\pm\)0.02 & 

& & & & & \\
\cmidrule{1-7}

\multicolumn{2}{l}{\textbf{Best F2-Scores}} 
& & Difference & -0.01 & P-Value & \cellcolor{red}0.3254 & 
& & & & & \\

\Xhline{0.8pt}  
\end{tabular}
\end{adjustbox}
\end{table*}

Among the evaluated models, GPT-4o, Claude 3.5 Haiku, and Gemini 1.5 Flash exhibited statistically significant improvements in F2-score performance when using prompts specifically engineered for each model, compared to the reused prompt P6. GPT-4o’s performance improved from 0.54 to 0.58 (p=0.0017). Claude 3.5 Haiku showed a substantial gain, with its F2-score increasing by 15\% (from 0.46 to 0.53, p=0.0001), while Gemini 1.5 Flash achieved the highest statistical significance, improving from 0.56 to 0.61 with a p-value of 0.0000. The results of Claude and Gemini were both statistically more significant and exhibited larger effect sizes than those observed for GPT-4o. This pattern may be attributed to the fact that the reused prompt was originally developed using GPT-4o-mini, a closely related model to GPT-4o. In contrast, Claude and Gemini differ in underlying architecture and training paradigms, making them more responsive to LLM-specific prompt optimization.

In contrast, several models did not exhibit statistically significant differences between the reused and LLM-specific prompts. Claude 3.5 Sonnet achieved nearly identical performance across both prompt types, with the re-used prompt yielding a slightly higher F2-score (0.58 vs. 0.57, p=0.1972). Gemini 1.5 Pro showed only a marginal improvement (+0.01, p=0.6528). Similarly, the open-source models LLaMA 3.1 70B and 8B demonstrated no significant gains; their best F2-scores remained stable or slightly declined, with p-values of 0.3254 and 0.5782, respectively. Notably, LLaMA 3.1 8B showed a sharp increase in F2-score under the 0-shot setting, from 0.07 with the reused prompt to 0.40 with the LLM-specific prompt. Upon inspecting the model’s predictions under the reused prompt, we found that it incorrectly assigned nearly all artifact pairs as not linked. This behavior appears to be triggered by the use of the word “directly” in the prompt, which may have led the model to overestimate the strength of relationships between artifacts. These results highlight that the reused prompt P6 already aligns reasonably well with the default behavior of these models, leaving little room for improvement. Additionally, the capabilities of some models are inherently limited, which may constrain their ability to benefit from more sophisticated or tailored prompts. Furthermore, the strong performance of P6 across multiple models highlights its generalizability and robustness as a baseline prompt in traceability tasks.

\vspace{5pt}
\noindent\fbox{
\parbox{0.95\columnwidth}{
\textbf{Observation 2:} While the re-used prompt (P6) demonstrates strong generalizability across models, LLM-specific prompt engineering yields statistically significant and practically meaningful improvements for certain models.
}
}
\vspace{5pt}

Across all models, 0-shot performance varies notably. For example, Gemini 1.5 Flash achieves the highest 0-shot F2-score (0.61), even surpassing its few-shot results, demonstrating its strong ability to handle traceability tasks without extra context. In contrast, some models struggle in 0-shot but improve with few-shots. For instance, LLaMA 3.1 8B model started with low 0-shot F2-score with the reused prompt (0.07) but significantly improved with just 2-shots (0.45). This demonstrates how few-shot learning can substantially enhance the model’s ability to handle the task, especially when its initial, demonstration-free performance is weak. Generally, most models show notable improvement in the few-shot settings. For instance, Gemini 1.5 Pro reaches the highest overall F2-score (0.58) with 2-shots, and GPT-4o achieves 0.58 with 6-shots. However, adding more demonstrations does not always enhance performance. For example, Gemini 1.5 Pro and Claude 3.5 Haiku perform better with 2-shots than with 4 or 6-shots.

\vspace{5pt}
\noindent\fbox{
    \parbox{0.95\columnwidth}{
        \textbf{Observation 3:} Few-shot learning with only two demonstrations is sufficient to enhance performance in most cases. Moreover, increasing the number of demonstrations does not always yield further improvements.
    }
}
\vspace{5pt}

Among all evaluated models, Gemini 1.5 Flash achieved the highest overall F2-score, reaching 0.61 in the zero-shot setting when paired with an LLM-specific prompt. Close behind were GPT-4o, GPT-4o-mini, and Gemini 1.5 Pro, each reaching a strong F2-score of 0.58, GPT-4o and GPT-4o-mini in 6-shot settings, and Gemini Pro in the 2-shot setting.  On the other hand, LLaMA models show lower performance, suggesting that they may not be as well-suited for traceability tasks in their current form. Interestingly, most lightweight models (Gemini Flash, GPT-4o-mini, and Claude Haiku) matched or outperformed their main counterparts (Gemini Pro, GPT-4o, and Claude Sonnet) despite having fewer parameters. These performance gains were especially pronounced when the lightweight models were used with carefully crafted, LLM-specific prompts, indicating that prompt engineering can play a significant role in utilizing their full potential. This indicates that smaller models, when paired with effective prompt engineering, can deliver comparable or superior performance in traceability tasks. 

\vspace{5pt}
\noindent\fbox{
\parbox{0.95\columnwidth}{
\textbf{Observation 4:} Lightweight models, when paired with LLM-specific prompts, can match or even outperform their larger counterparts, demonstrating that effective prompt engineering can compensate for reduced model capacity in traceability tasks.
}
}
\vspace{5pt}

\subsection{RQ3: To what extent do demonstrations selection strategies influence few-shot learning performance?}

This section answers RQ3 by conducting experiments with each selection strategy using both label-aware sampling (i.e., applying a balanced label constraint) and the standard unconstrained approach. Our results indicate that applying the label-aware sampling strategy consistently improves performance across all datasets and numbers of shots, regardless of the selection strategy used. These findings are in line with existing work~\citep{fei-etal-2023-mitigating, gao2025exploring, du_automatic_2020}. Given these findings, we present the results using the label-aware sampling constraint in the following sections. 

The best-performing model identified in RQ2 is Gemini 1.5 Flash under the 0-shot setting. However, since this setting does not involve demonstrations, we considered the next best-performing models: GPT-4o-Mini, GPT-4o, and Gemini 1.5 Pro. For this experiment, we used GPT-4o-mini due to its superior efficiency compared to the others.

Table~\ref{tab:DSS_results} presents a comparison of the effect of the demonstrations selection strategies on performance across different shot settings, using the Random strategy as a baseline. The analysis reveals that different demonstrations selection strategies have varying impacts on performance. In terms of precision, the Similarity-based strategy consistently performs best, peaking at 0.52 in the 6-shot setting. In contrast, Diversity-based and Uncertainty-based strategies yield lower precision, with the latter dropping notably at higher shots. Regarding recall, the Diversity strategy outperforms others at 2-shot (0.82) but experiences a decline at higher shots, while the Random strategy maintains relatively stable recall across different shots. For F2-score, which emphasizes recall, the Diversity strategy achieves the best performance at 2-shot (0.68), while Random, Uncertainty, and Similarity-based selections converge to similar values. 

Importantly, the statistical significance of the differences in F2-score between each strategy and the Random baseline was evaluated using p-values. The results show that Diversity-based selection yields a statistically significant improvement (p = 0.000) over Random, indicating a reliable performance gain. In contrast, Similarity-based (p = 0.182) and Uncertainty-based (p = 0.264) strategies do not show statistically significant improvements at the 0.05 level. This suggests that Diversity-based selection with only 2 shots is the most effective approach for improving recall and F2-score, making it the preferred choice in traceability scenarios, where recall is critical.

\begin{table}[]
\centering
\caption{A comparison of DSS effect on GPT-4o-Mini performance across different shots.}
\label{tab:DSS_results}

    \begin{tabular}{@{}lcccc@{}}
    \toprule
    \textbf{Shots}  & \textbf{Random} & \textbf{Diversity} & \textbf{Similarity} & \textbf{Uncertainty} \\ 
    \midrule

    \multicolumn{5}{c}{\textbf{Precision}} \\ \midrule
    2-shot  & 0.42\(\pm\)0.04 & 0.40\(\pm\)0.02 & 0.51\(\pm\)0.03 & 0.49\(\pm\)0.03 \\ 
    4-shot  & 0.46\(\pm\)0.06 & 0.47\(\pm\)0.03 & 0.48\(\pm\)0.03 & 0.43\(\pm\)0.02 \\ 
    6-shot  & 0.46\(\pm\)0.05 & 0.48\(\pm\)0.04 & 0.52\(\pm\)0.02 & 0.37\(\pm\)0.03 \\ 
    \midrule
    \multicolumn{5}{c}{\textbf{Recall}} \\ \midrule
    2-shot  & 0.58\(\pm\)0.06 & 0.82\(\pm\)0.03 & 0.44\(\pm\)0.00 & 0.52\(\pm\)0.05 \\ 
    4-shot  & 0.58\(\pm\)0.08 & 0.57\(\pm\)0.03 & 0.59\(\pm\)0.03 & 0.66\(\pm\)0.05 \\ 
    6-shot  & 0.62\(\pm\)0.03 & 0.56\(\pm\)0.05 & 0.60\(\pm\)0.02 & 0.62\(\pm\)0.05 \\ 
    \midrule
    \multicolumn{5}{c}{\textbf{F2-score}} \\ \midrule
    2-shot  & 0.53\(\pm\)0.04 & \textbf{0.68\(\pm\)0.02} & 0.51\(\pm\)0.02 & 0.51\(\pm\)0.04 \\ 
    4-shot  & 0.55\(\pm\)0.05 & 0.54\(\pm\)0.03 & 0.56\(\pm\)0.03 & 0.59\(\pm\)0.04 \\ 
    6-shot  & 0.58\(\pm\)0.03 & 0.54\(\pm\)0.04 & 0.58\(\pm\)0.02 & 0.55\(\pm\)0.04 \\ 
    \midrule
    \multicolumn{2}{l}{\textbf{P-Value (F2-Score)}}  & \cellcolor{green}0.000 & \cellcolor{red}0.182 & \cellcolor{red}0.264 \\ 
    \bottomrule
    \end{tabular}
\end{table}

\vspace{5pt}
\noindent\fbox{
\parbox{0.95\columnwidth}{
\textbf{Observation 5:} Label-aware sampling consistently improves performance across models, shots, and selection strategies. Diversity-based selection with only 2 shots yields the highest recall and F2-score with a significant statistical difference, making it the preferred choice for traceability tasks.
}
}
\vspace{5pt}

\subsection{RQ4: To what extent does the proposed prompt-engineered LLM approach generalize across domains (e.g., healthcare, aerospace) and artifact types (e.g., requirements, test cases), compared to existing baselines and state-of-the-art traceability methods?}\label{sec:RQ4}

This section answers RQ4 by comparing the performance of TraceLLM on different datasets to the baselines and state-of-the-art methods. In order to adapt the designed prompt (P6) to different datasets: EasyClinic (UC-TC), EasyClinic (UC-ID), and CCHIT, we converted the prompt to a template to be adjusted to dataset-specific details (see Table~\ref{tab:p6_template_values}), as follows: 

\textit{``You are an expert in software traceability. You are given two artifacts from [DOMAIN] system. (1) is [ARTIFACT\_1] and (2) is [ARTIFACT\_2]. Does [RELATION]? Answer with only `Yes' or `No'.''}

\begin{table}[]
\centering
\small
\caption{Values used to populate the P6 prompt template for each dataset.}
\label{tab:p6_template_values}
\begin{tabular}{@{}lllll@{}}
\toprule
\textbf{Dataset} & \textbf{DOMAIN} & \textbf{ARTIFACT\_1} & \textbf{ARTIFACT\_2} & \textbf{RELATION} \\
\midrule
CM1 & Aerospace & High-level requirement & Design element & (2) fulfill (1) \\
EasyClinic & Healthcare & Use case & Test case & (2) test (1) \\
EasyClinic & Healthcare & Use case & Interaction Diagram & (2) realize (1) \\
CCHIT & Healthcare & Requirement & Regulation & (1) satisfy (2) \\
\bottomrule
\end{tabular}
\end{table}

The results, summarized in Table~\ref{tab:baselines_comparison}, show that TraceLLM consistently achieves the highest F2-scores across most datasets. On EasyClinic (UC-TC), TraceLLM outperforms all methods with an F2-score of 0.83, substantially higher than \citet{hey2025requirements} (0.72) and IR baselines such as VSM (0.66) and LSI (0.65). This indicates that TraceLLM not only shows high recall (0.93) but also improves precision (0.58), yielding a more balanced and effective traceability. A similar trend is observed in EasyClinic (UC-ID), where TraceLLM again dominates with an F2-score of 0.82, outperforming all methods. For example, TraceLLM outperforms both \citet{hey2025requirements} (0.63) and \citet{rodriguez_prompts_2023}. Notably, TraceLLM’s precision on this dataset is especially high (0.89), suggesting that it is capable of retrieving highly relevant links even in settings with complex artifact structures like interaction diagrams.

In the CCHIT dataset, which evaluates the traceability between requirements and regulations, TraceLLM achieves an F2-score of 0.69, outperforming all baselines and state-of-the-art methods. This result is particularly significant because \citet{etezadi2025classification} explicitly designed a handcrafted prompt for this exact artifacts pair. Despite this, TraceLLM substantially surpasses \citet{etezadi2025classification}’s method (0.30). Furthermore, TraceLLM outperforms \citet{hey2025requirements} on this dataset (0.29), and it exceeds BERT by a large margin (0.16).

The CM1 dataset presents a slightly more competitive comparison. \citet{hey2025requirements} achieves the highest F2-score (0.69), with TraceLLM closely behind at 0.68. While the difference is marginal, TraceLLM maintains high recall (0.82) and competitive precision (0.40).

Taken together, these results offer strong evidence that prompt-engineered LLMs, as implemented in TraceLLM, are capable of generalizing across domains and artifact types without the need for dataset-specific customization or prompt-tuning. Moreover, the superior F2-scores achieved by TraceLLM in three out of four datasets, combined with consistently high recall, affirm its suitability for traceability scenarios where recall is a priority, such as safety-critical or compliance-driven environments.

\vspace{5pt}
\noindent\fbox{
\parbox{0.95\columnwidth}{
\textbf{Observation 6:} TraceLLM generalizes effectively across domains and artifact types, achieving the highest F2-score on most datasets. These results highlight TraceLLM’s robustness and suitability for recall-critical traceability scenarios.
}
}
\vspace{5pt}

To better understand the impact of the selected few-shot examples on TraceLLM’s performance, we evaluated the framework under zero-shot and 2-shot random settings across all four datasets. As shown in Table~\ref{tab:baselines_comparison}, TraceLLM outperforms traditional IR baselines and BERT-based methods across all datasets, and imperove over the prior LLM-based approaches in three out of four datasets even in the zero-shot setting. Introducing two randomly selected demonstrations does not consistently improve performance compared to the zero-shot setting and, in several cases, results in lower F2-scores. Moreover, random few-shot selection consistently underperforms label-aware, diversity-based demonstration selection, highlighting that the benefit of few-shot learning depends strongly on how demonstrations are selected rather than on their presence. Overall, these findings suggest that TraceLLM’s gains are not solely driven by the availability of labeled examples, but that systematic demonstration selection remains important for maximizing performance.

\begin{sidewaystable}[]
\centering
\caption{Comparison of TraceLLM with baselines and state-of-the-art methods across four datasets.}
\small
\label{tab:baselines_comparison}
\begin{tabular}{@{}lllllll@{}}
\toprule

\multirow{2}{*}{\textbf{Model}} &
\multicolumn{3}{l}{\textbf{CM1}} &
\multicolumn{3}{l}{\textbf{EasyClinic (UC-TC)}} \\
\cmidrule{2-4} \cmidrule{5-7}
& \textbf{Precision} & \textbf{Recall} & \textbf{F2-Score} & \textbf{Precision} & \textbf{Recall} & \textbf{F2-Score} \\
\midrule
VSM & 0.32±0.00 & 0.61±0.00 & 0.52±0.00 & 0.49±0.00 & 0.72±0.00 & 0.66±0.00 \\ 
LSI & 0.28±0.02 & 0.67±0.07 & 0.52±0.03 & 0.48±0.09 & 0.73±0.13 & 0.65±0.05 \\ 
LDA & 0.26±0.04 & 0.54±0.07 & 0.45±0.05 & 0.40±0.09 & 0.86±0.14 & 0.70±0.12 \\
BERT & 0.67±0.00 & 0.22±0.00 & 0.26±0.00 & 0.16±0.00 & 0.80±0.00 & 0.44±0.00 \\ 
\citet{rodriguez_prompts_2023} & 0.06±0.00 & 1.00±0.00 & 0.23±0.00 & 0.25\(\pm\)0.00 & 0.96\(\pm\)0.00 & 0.62\(\pm\)0.00 \\ 
\citet{etezadi2025classification} & - & - & - & - & - & - \\ 
\citet{hey2025requirements} & 0.44±0.02 & 0.80±0.03 & \textbf{0.69±0.02} & 0.34±0.02 & 1.00±0.00 & 0.72±0.01 \\
\midrule
TraceLLM & 0.40\(\pm\)0.02 & 0.82\(\pm\)0.03 & 0.68\(\pm\)0.02 & 0.58\(\pm\)0.02 & 0.93\(\pm\)0.02 & \textbf{0.83\(\pm\)0.01} \\
TraceLLM (0-shot) & 0.52\(\pm\)0.03 & 0.59\(\pm\)0.03 & 0.57\(\pm\)0.03 & 0.72\(\pm\)0.03 & 0.80\(\pm\)0.04 & 0.79\(\pm\)0.04 \\
TraceLLM (2-shot random) & 0.42\(\pm\)0.04 & 0.58\(\pm\)0.06 & 0.53\(\pm\)0.04 & 0.63\(\pm\)0.07 & 0.79\(\pm\)0.14 & 0.75\(\pm\)0.12 \\
\toprule

\multirow{2}{*}{\textbf{Model}} &
\multicolumn{3}{l}{\textbf{EasyClinic (UC-ID)}} &
\multicolumn{3}{l}{\textbf{CCHIT}} \\
\cmidrule{2-4} \cmidrule{5-7}
& \textbf{Precision} & \textbf{Recall} & \textbf{F2-Score} & \textbf{Precision} & \textbf{Recall} & \textbf{F2-Score} \\
\midrule
VSM & 0.44±0.00 & 0.70±0.00 & 0.63±0.00 & 0.12±0.00 & 0.39±0.00 & 0.26±0.00 \\
LSI & 0.40±0.03 & 0.54±0.15 & 0.49±0.09 & 0.10±0.01 & 0.45±0.04 & 0.26±0.01 \\
LDA & 0.37±0.05 & 0.52±0.20 & 0.47±0.13 & 0.03±0.02 & 0.13±0.08 & 0.06±0.02 \\
BERT & 0.11±0.00 & 0.60±0.00 & 0.32±0.00 & 0.05±0.00 & 0.29±0.00 & 0.16±0.00 \\
\citet{rodriguez_prompts_2023} & 0.19±0.00 & 1.00±0.00 & 0.54±0.00 & 0.04±0.00 & 0.94±0.00 & 0.15±0.00 \\
\citet{etezadi2025classification} & - & - & - & 0.08±0.00 & 0.81±0.00 & 0.30±0.00 \\ 
\citet{hey2025requirements} & 0.26±0.01 & 1.00±0.00 & 0.63±0.02 & 0.10±0.00 & 0.57±0.00 & 0.29±0.00 \\
\midrule
TraceLLM & 0.89±0.00 & 0.80±0.00 & 0.82±0.00 & 0.41±0.01 & 0.84±0.00 & \textbf{0.69±0.01} \\
TraceLLM (0-shot) & 0.82\(\pm\)0.00 & 0.90\(\pm\)0.00 & \textbf{0.88\(\pm\)0.00} & 0.33\(\pm\)0.01 & 0.79\(\pm\)0.02 & 0.62\(\pm\)0.01 \\
TraceLLM (2-shot random) & 0.79\(\pm\)0.06 & 0.82\(\pm\)0.10 & 0.81\(\pm\)0.08 & 0.23\(\pm\)0.03 & 0.89\(\pm\)0.01 & 0.57\(\pm\)0.03 \\
\bottomrule
\end{tabular}
\end{sidewaystable}

\begin{figure}
    \centering
    \includegraphics[width=\linewidth]{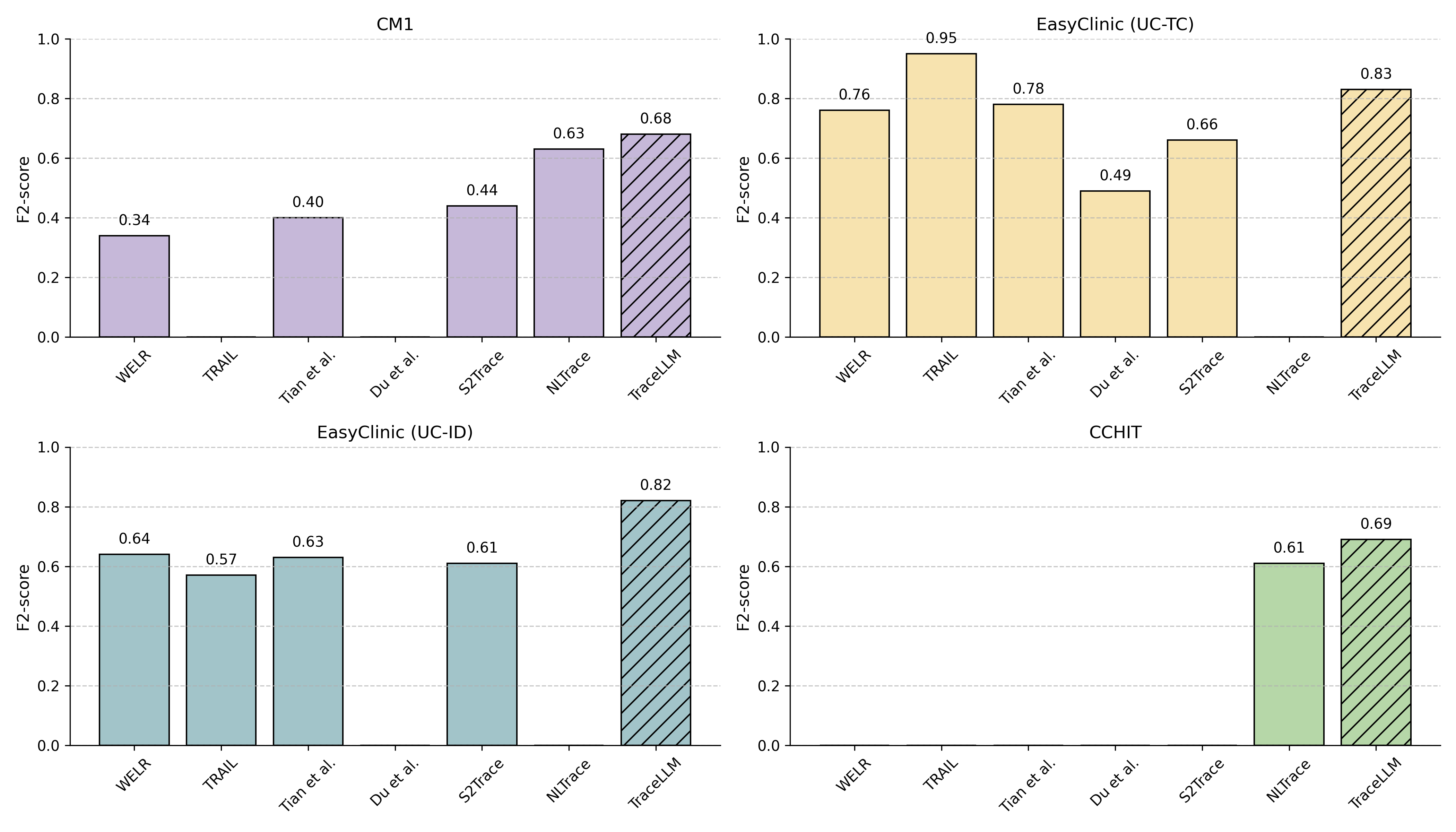}
    \caption{F2-score qualitative comparison of TraceLLM with state-of-the-art traceability approaches across four datasets. Absence of a bar indicates that the method was not evaluated on the corresponding dataset.}
    \label{fig:comparison_ml_dl_lm}
\end{figure}

To complement our quantitative evaluation, we conducted a qualitative comparison with recent ML, DL, and language models (e.g., BERT) traceability approaches that used similar datasets but did not release their data partitions or replication packages. As shown in Figure~\ref{fig:comparison_ml_dl_lm}, TraceLLM consistently outperforms all prior methods across most datasets where comparisons are possible, highlighting its effectiveness and generalizability.

On the CM1 dataset, TraceLLM achieves an F2-score of 0.68, outperforming all methods, such as NLTrace (0.63), S2Trace (0.44), and \citet{tian_adapting_2018} (0.40). For EasyClinic (UC-TC) and (UC-ID), TraceLLM achieves F2-scores of 0.83 and 0.82, respectively, surpassing established baselines such as WELR (0.76 and 0.64), \citep{tian_adapting_2018} (0.78 and 0.63). While TRAIL reports a high F2-score (0.95) on UC-TC, it underperforms significantly on UC-ID (0.57), suggesting that its performance may be sensitive to specific artifacts. In contrast, TraceLLM delivers strong, balanced performance across all artifact types, reflecting greater robustness. In the CCHIT dataset, TraceLLM attains the highest reported F2-score of 0.69, exceeding NLTrace (0.61).

Overall, the qualitative comparison highlights that TraceLLM not only matches but often exceeds the performance of prior methods, even those designed with domain-specific tuning or handcrafted features. Unlike existing ML and DL approaches, which typically require extensive training, feature engineering, or dataset-specific adaptation, TraceLLM operates in a zero- or few-shot setting using a general-purpose prompt. 

\vspace{5pt}
\noindent\fbox{
\parbox{0.95\columnwidth}{
\textbf{Observation 7:} TraceLLM outperforms prior ML, DL, and language model-based methods across multiple datasets, despite requiring no training or handcrafted features. This highlights the effectiveness of prompt-engineered LLMs as a general-purpose solution for software traceability.
}
}
\vspace{5pt}

\subsubsection{Effect of Dataset-Specific Prompt Optimization}

In addition to evaluating cross-dataset generalization, we performed preliminary experiment to investigate whether dataset-specific prompt optimization leads to significant performance improvements compared to reusing a prompt optimized on a different dataset. This analysis aims to better understand the practical benefit of per-project prompt optimization for practitioners.

Specifically, we performed prompt optimization on the EasyClinic\_UC\_TC dataset following the same TraceLLM procedure. The resulting dataset-specific prompt was:
\textit{``(1) is a use case and (2) is a test case. Does (2) test (1)? Answer with only `Yes' or `No'."}. We then compared the performance of this dataset-specific prompt against the performance obtained on EasyClinic\_UC\_TC using the prompt optimized on CM1\_NASA.

The results show that dataset-specific prompt optimization yields similar performance to prompt reuse. Using the dataset-specific prompt, TraceLLM achieved an average precision of 0.60\(\pm\)0.02, recall of 0.92\(\pm\)0.00, and F2-score of 0.83\(\pm\)0.01, compared to 0.58\(\pm\)0.02 precision, 0.93\(\pm\)0.02 recall, and 0.83\(\pm\)0.01 F2-score when using the prompt optimized on CM1. This comparison indicates that prompt optimization on the target dataset provides limited additional benefit in this case.

From a practical perspective, this finding suggests that TraceLLM’s prompts generalize well across datasets, and that practitioners may achieve competitive performance without incurring the additional cost and effort of dataset-specific prompt optimization. While project-specific tuning may still be beneficial in some settings, these results indicate that prompt reuse is a viable and efficient option when labeled data or optimization resources are limited.

\subsection{Cost Analysis}

We provide a cost analysis in Table~\ref{tab:cost_comparison}, summarizing the total expenses for input/output tokens across 30 experimental runs, which include 5 for zero-shot configuration and 25 runs for few-shot settings (5 repetitions over 5 demonstrations sets) on the CM1 test set. Lightweight models such as GPT-4o-Mini (\$5.909) and Gemini 1.5 Flash (\$2.955) demonstrate excellent cost-efficiency. In contrast, flagship models like Claude 3.5 Sonnet (\$118.228), GPT-4o (\$98.485), and Gemini 1.5 Pro (\$49.243) require substantially higher costs without performance gains. Among the options, GPT-4o-Mini stands out as a particularly affordable yet effective model, achieving a strong F2-score of 0.69 while maintaining low costs.

These findings are particularly relevant for both academic researchers and software practitioners seeking scalable, cost-effective solutions for automated traceability. For researchers, they highlight the viability of lightweight LLMs in experimental setups without sacrificing quality. For practitioners, especially those operating under budgetary constraints or deploying models in production environments, these models offer a compelling balance between cost and performance, supporting practical adoption of LLM-based traceability in real-world software engineering workflows.

\begin{table}[h!]
\centering
\caption{Cost comparison of different LLMs.}
\small
\label{tab:cost_comparison}
\begin{tabular}{@{}llll@{}}
\toprule
\textbf{Model} & \multicolumn{3}{l}{\textbf{Cost in US \$}} \\ \cmidrule(lr){2-4} 
               & \textbf{Input} & \textbf{Output} & \textbf{Total} \\ \midrule
GPT-4o-Mini    & 5.900          & 0.009           & 5.909         \\
GPT-4o         & 98.334        & 0.151           & 98.485        \\
Claude 3.5 Haiku & 9.833        & 0.019           & 9.852        \\
Claude 3.5 Sonnet & 118.001      & 0.227           & 118.228        \\
Gemini 1.5 Flash & 2.950        & 0.005           & 2.955         \\
Gemini 1.5 Pro & 49.167          & 0.076           & 49.243         \\
LLAMA 3.1 8B Turbo* & 7.080      & 0.003           & 7.083         \\
LLAMA 3.1 70B Turbo* & 34.614     & 0.013           & 34.627         \\ \bottomrule
\multicolumn{4}{l}{\footnotesize *Using Together inference API \url{https://www.together.ai}} \\
\end{tabular}
\end{table}

\subsection{Implications for Practice}\label{sec:practice}

The traceability literature consistently indicates that fully automated trace recovery remains an open challenge, as current techniques do not yet achieve the level of performance required for uncritical use in large-scale systems~\citep{hayes2003improving, clelandhuang2007,delucia2008}. Prior work across information retrieval, machine learning, and more recent LLM-based approaches reports performance levels that fall short of enabling fully automated traceability~\cite{guo2017semantically, hayes2003improving, rodriguez_prompts_2023, hey2025requirements, etezadi2025classification}. While TraceLLM improves performance relative to existing baselines, achieving recall values in the range of approximately 60--70\% and corresponding F2-scores around 0.6--0.7, these results are still not sufficient to support fully automated traceability without human oversight.

A key reason for the continued difficulty of full automation lies in the trade-off between precision and recall, and in identifying practitioner expectations regarding acceptable traceability quality. Some practitioners emphasize precision, expressing strong concerns about incorrect trace links that may undermine trust or introduce risks, particularly in safety-critical or compliance-driven contexts~\citep{ramesh2001toward,aizenbud2006}. Other studies highlight that missing true links (false negatives) can be equally or more costly, as they may lead to overlooked change impacts, incomplete test coverage, or missed compliance obligations~\citep{lin2022enhancing, lin2022information, hayes2006advancing}. This tension has led traceability research to frequently emphasize recall by using metrics such as F2-score, under the assumption that automated techniques are used to support human analysts rather than replace them.

In this context, TraceLLM is best understood as a semi-automated, decision-support approach. Its practical contribution lies in improving recall relative to existing approaches, thereby helping analysts identify candidate trace links that might otherwise be missed. False positives generated by TraceLLM can be discarded during validation, while the improved recall reduces the risk of missing relevant links. Rather than delivering final traceability decisions, TraceLLM reduces the analyst search space and supports trace link completion and recovery tasks. This positioning is consistent with both classical trace recovery research and recent LLM-based traceability studies, and reflects realistic usage scenarios in practice~\citep{cleland2012software}.

\section{Threats to Validity}\label{sec:TTV}

As with empirical studies, our findings are subject to several threats to validity. In this section, we discuss the main threats and describe the steps taken to mitigate their impact where possible.

\subsection{External Validity}
The prompt design and tuning process was initially performed on the CM1 dataset, which may introduce dataset-specific biases and limit generalizability. To mitigate this threat, we evaluated the finalized prompts on three additional publicly available benchmark datasets spanning different domains and artifact types, including healthcare and regulatory documents. While these datasets are widely used in prior traceability research and vary in complexity, size, and artifact characteristics, they may not fully represent the diversity and scale of real-world industrial software projects. As a result, the generalizability of the findings to other domains, artifact types, or large-scale industrial settings may be limited.

\subsection{Construct Validity}
Our evaluation emphasizes recall, particularly using the F2-score metric, reflecting common assumptions in trace recovery research where missing true links can be costly. However, this choice implicitly assumes human-in-the-loop validation, where false positives can be filtered during review. Although TraceLLM improves recall compared to existing approaches, the achieved performance levels are not sufficient for fully automated traceability, specifically in large-scale or safety-critical systems. In workflows that require fully automated traceability with minimal tolerance for incorrect links, this evaluation focus may not align with practitioner expectations. A As with prior automated trace recovery techniques~\citep{rodriguez_prompts_2023, hassine2024llm, lin2022enhancing}, TraceLLM requires human validation and is therefore best positioned as a semi-automated, decision-support approach that supports practitioners in identifying trace links. To mitigate these threats, we explicitly positioned TraceLLM as a semi-automated, decision-support approach and discussed the practical implications of these assumptions in Section~\ref{sec:practice}. 

In addition, while TraceLLM supports both zero-shot and few-shot settings, its strongest performance is achieved in few-shot configurations that rely on a small number of labeled trace links. This limits applicability in early-stage or cold-start projects with no existing traceability information, making TraceLLM more effective for trace link completion (TLC) than for trace link generation (TLG) tasks. Nevertheless, we additionally report zero-shot and random few-shot results across all datasets; notably, TraceLLM still outperforms the baselines under these settings, indicating that its effectiveness is not solely driven by the selected demonstrations.

\subsection{Conclusion Validity}
As with other machine learning research, TraceLLM does not guarantee globally optimal prompts~\citep{reynolds2021prompt}. While the methodology provides a structured and repeatable procedure for prompt refinement, unexplored prompt variants may yield additional performance gains. The prompt engineering process, despite being systematic, remains sensitive to phrasing and involves subjective decisions. Minor wording changes (e.g., using “directly fulfill” instead of “fulfill”) had notable effects on model behavior. To reduce this subjectivity, we conducted iterative refinement and comparative evaluations of prompt variants on the validation set and incorporated role-specific and domain-specific enrichment. Our conclusions, therefore, hold under comparable prompt engineering effort rather than assuming prompt optimality. 

In addition, LLM outputs are inherently non-deterministic and dependent on underlying model implementations and infrastructure. To mitigate this threat, each experiment was repeated multiple times and results are reported using mean and standard deviation. Nevertheless, minor variability may persist due to factors beyond the authors’ control, such as backend updates to proprietary models.

\section{Conclusion and Future Work}\label{sec:conclusion}
This study presented TraceLLM, a systematic framework that leverages prompt engineering with LLMs to enhance automated software traceability. By formulating and refining prompts through a structured, iterative process and evaluating them across multiple LLMs and datasets, we demonstrated that LLMs, particularly in few-shot settings, can achieve superior performance compared to traditional and state-of-the-art methods. The proposed approach achieved strong F2-scores across diverse traceability tasks and domains, highlighting the potential of prompt-engineered LLMs to serve as a general-purpose solution for traceability scenarios where recall is critical.

Furthermore, we explored the impact of demonstrations selection strategies in few-shot learning, identifying diversity-based and label-aware sampling as particularly effective. Our findings also revealed that lightweight models, when paired with carefully designed prompts, can match or even exceed the performance of larger models, offering a cost-effective alternative for practical deployment. 

Future work may involve evaluating TraceLLM in the context of trace link maintenance and evolution within dynamic project environments, exploring advanced prompting techniques such as Chain-of-Thought (CoT) and Tree-of-Thought (ToT), and assessing the benefits of fine-tuning LLMs for domain-specific traceability tasks. Additionally, expanding the evaluation to industrial datasets and a wider range of artifact types could offer further insights into the scalability and adaptability of LLM-based traceability approaches.

\appendix

\begin{landscape}
\section{Baseline Configurations}
\label{app:baselines_configurations}

\begin{table}[h!]
\centering
\caption{Summary of baseline LLM-based traceability approaches: best reported model, shot settings, and prompt.}
\label{tab:baseline_prompts}
\begin{tabular}{p{2.5cm} p{3cm} p{1.5cm} p{10cm}}
\toprule
\textbf{Study} & \textbf{Model(s)} & \textbf{\#Shots} & \textbf{Prompt} \\
\midrule
\citet{rodriguez_prompts_2023} & \textit{claude-instant-v1} & Zero-shot & I am giving you two software artifacts from a system. Your job is to determine if there is a traceability link. 
Answer whether (2) [RELATION] a part of (1) with yes or no enclosed in \textless[RELATION]\textgreater \textless/[RELATION]\textgreater. 
Use your answers to give one reason why (1) might be related to (2) enclosed in \textless related\textgreater \textless/related\textgreater and one reason why (1) might be un-related to (2) enclosed in \textless unrelated\textgreater \textless/unrelated\textgreater 
Now answer is (1) related to (2) with yes or no enclosed in \textless traced\textgreater \textless/traced\textgreater.'' \\
\midrule
\citet{hey2025requirements} & Embedding model: \textit{text-embedding-3-large} Prompting model: \textit{gpt-4o} & Zero-shot & Below are two artifacts from the same software system. Is there a traceability link between (1) and (2)? Give your reasoning and then answer with 'yes' or 'no' enclosed in \textless trace\textgreater \textless/trace\textgreater.\\
\midrule

\citet{etezadi2025classification} & \textit{gpt-4o} & Five-shots & 
\textbf{[Context]} I am currently working on a task focused on establishing traceability between software requirements and regulatory codes. This involves analyzing and mapping requirements to relevant HIPAA regulations, ensuring that our software development aligns with regulatory compliance. 
Below are the main regulatory codes that I want you to remember at first: AC: Access Control. Implement technical policies... etc. 

\textbf{[Examples]} Here are five sample traceability examples. I've also added my rationale for tracing regulatory codes to the requirements for your reference. 
Requirement: The system shall provide the ability for authorized administrators to assign restrictions or privileges to users/groups.
Trace links: [AC] 
Rational behind choosing these codes: The requirement supports... etc.

\textbf{[Instructions]} Find the trace links for a given requirement and provide the rationale behind your choice extended from the examples I provided. Please consider regulatory codes which I have not used in the examples. Pay attention to the roles (AS\_ROLE) in the requirement, if there is any. Remember, regulations text focus on personal data, but try to consider all types of data, role, or functionalities in a software system. Pay attention to commonsense and indirect relations between requirement and regulations. Aim to include regulations even if they have a low likelihood of being traced, prioritizing recall over precision. Choose at least one regulation for each requirement.

\textbf{[Output]}List of alphabetical order of regulatory codes (if any) similar to the examples I provided to you. Newline to explain the rational behind the choice(s).\\
\bottomrule
\end{tabular}
\end{table}
\end{landscape}






\bibliography{sn-bibliography}

\end{document}